# A.-M. Guerry's Moral Statistics of France: Challenges for Multivariable Spatial Analysis

**Michael Friendly**

*Abstract.* André-Michel Guerry's (1833) *Essai sur la Statistique Morale de la France* was one of the foundation studies of modern social science. Guerry assembled data on crimes, suicides, literacy and other "moral statistics," and used tables and maps to analyze a variety of social issues in perhaps the first comprehensive study *relating* such variables. Indeed, the *Essai* may be considered the book that launched modern empirical social science, for the questions raised and the methods Guerry developed to try to answer them.

Guerry's data consist of a large number of variables recorded for each of the départments of France in the 1820–1830s and therefore involve both multivariate and geographical aspects. In addition to historical interest, these data provide the opportunity to ask how modern methods of statistics, graphics, thematic cartography and geovisualization can shed further light on the questions he raised. We present a variety of methods attempting to address Guerry's challenge for multivariate spatial statistics.

## 1. INTRODUCTION

On July 2, 1832 a slim manuscript was presented to the Académie Française des Sciences by the 29 year old lawyer André-Michel Guerry titled *Essai sur la Statistique Morale de la France*. Guerry's findings were both startling and compelling. His presentation, in tables and *cartes figuratives*, of statistical data on crime, suicide and other moral aspects, measured only recently in France, broke new ground in thematic cartography and data visualization. Along with the nearly simultaneous work of Adolphe Quetelet (Quetelet (1831), 1835) in Belgium, Guerry's *Essai* (published in 1833) established the scientific study of "moral statistics" in Europe and became the launching pad for much of modern social science: criminology and sociology in particular.

Guerry's results were startling for two reasons. First he showed that rates of crime and suicide remained remarkably stable over time, when broken down by age, sex, region of France and even season of the year; yet these numbers varied systematically across départements of France. This regularity of social numbers created the possibility to conceive, for the first time, that human actions in the social world were governed by social laws, just as inanimate objects were governed by laws of the physical world. By extension, these laws could be uncovered by the careful collection and analysis of social "facts," meaning numbers. Second, he overturned some widespread beliefs about the nature and causes of crime and its relation to other observable factors, such as education and poverty. Over his lifetime, he completed three major works on moral statistics, winning the Montyon prize in statistics twice from the Académie Française des Sciences. His last work (1864) contemplated multivariate explanations of relations among moral variables, at a time well before the development of correlation and regression. Yet Guerry's contributions to statistics, graphics and the rise of modern social science in the early 1800s are neither well known nor widely appreciated outside criminology and sociology.

*Michael Friendly is Professor, Psychology Department, York University, Toronto, Ontario, M3J 1P3 Canada e-mail: friendly@yorku.ca.*

This paper recounts Guerry's work, the questions he asked and the methods he used to answer them in relation to his place in the history of data visualization and statistics. To do so, we first describe the context in which he worked and which led to the rise of the moral statistics movement in Europe. A second section describes his life and works and the methods he introduced to the study of moral statistics. Guerry worked with voluminous data on social variables, distributed over time and space (départements of France, counties of England), and finely categorized along numerous dimensions (age, sex and status of accused, detailed breakdown of types of personal and property crime, motives for suicide, etc.). The final sections attempt some reanalyses of Guerry's data to address the challenges posed for modern multivariate and spatial analysis.

### 1.1 The Rise of "Moral Statistics" and Modern Social Science

The empirical and quantitative study of factors affecting human society such as education, crime and poverty that gave rise to modern social science began between 1825 and 1835, with the work of André-Michel Guerry and Adolphe Quetelet. But the roots of this endeavor and the very possibility of observation-based laws governing human populations go back much further.

The systematic study of social numbers, at first concerning population data and dynamics, began in the 1660s with John Graunt's (1662) and William Petty's (1665) analyses of the London *Bills of Mortality*. This work showed how such numbers could inform the state about matters related to population growth, age-specific mortality, ability to raise an army, the consequences of plagues, wealth, taxes and so forth. "Political arithmetic," as it was called, was based only on the simple ideas of standardizing raw numbers by relevant totals and the "rule of three," $a : b$ as $c : ?$, to make proportional comparisons; but by these means political arithmeticians were able to establish a basis for comparisons over geographic region, time, age and other categories (Klein (1997)).

The *Bills of Mortality* were based on parish records of christenings and deaths, recorded nearly weekly and with at least a modicum of uniformity. In 1710, John Arbuthnot (1710), a Scottish minister and physician to Queen Anne, calculated the ratio of male to female births from these records for 1629–1710 and observed that the ratio was consistently greater than 1 (see Figure 1). He used this lawful regularity to argue that divine providence, not chance, governs the sex ratio, in perhaps the first application of probability to social statistics and the first formal significance test.

By the mid-1700s, the importance of measuring and analyzing population distributions and the idea that ethical and state policies could encourage wealth through population growth was established, most notably by Johann Peter Süssmilch (1741), who advocated expansion of governmental collection of population statistics. Data on the social character of human populations was still lacking, however.

In the period leading up to and through the Bourbon Restoration following Napoleon's defeat in 1815, crime was a serious concern, particularly in Paris, which witnessed an explosive growth in population, along with widespread inflation and unemployment, and the emergence of an impoverished, dangerous class of petty criminals (les misérables); see Beirne (1993a) and Chevalier (1958). Then, as now, there were two basic schools of thought regarding criminal justice policy and much debate about the treatment of prisoners. A liberal, *philanthrope* position advocated increased education, religious instruction, improved diet (bread *and soup*!) and better prison conditions as the means to reduce crime and recidivism. Hard-line, conservatives feared attempts at prison reform, doubted the efficacy of campaigns for public education and viewed suggestions to abandon the harsh punishment of convicts under the *ancien régime* with grave suspicion if not alarm. But the evidence marshalled to support such recommendations was fragmentary, restricted and often idiosyncratic. See Whitt (2002, pages xxvi–xxxi), Beirne (1993b) and Porter (1986, pages 27–30) for more background on the social context in which the *Essai* was written.

This changed in 1825 when the Ministry of Justice in France instituted the first centralized, *national*

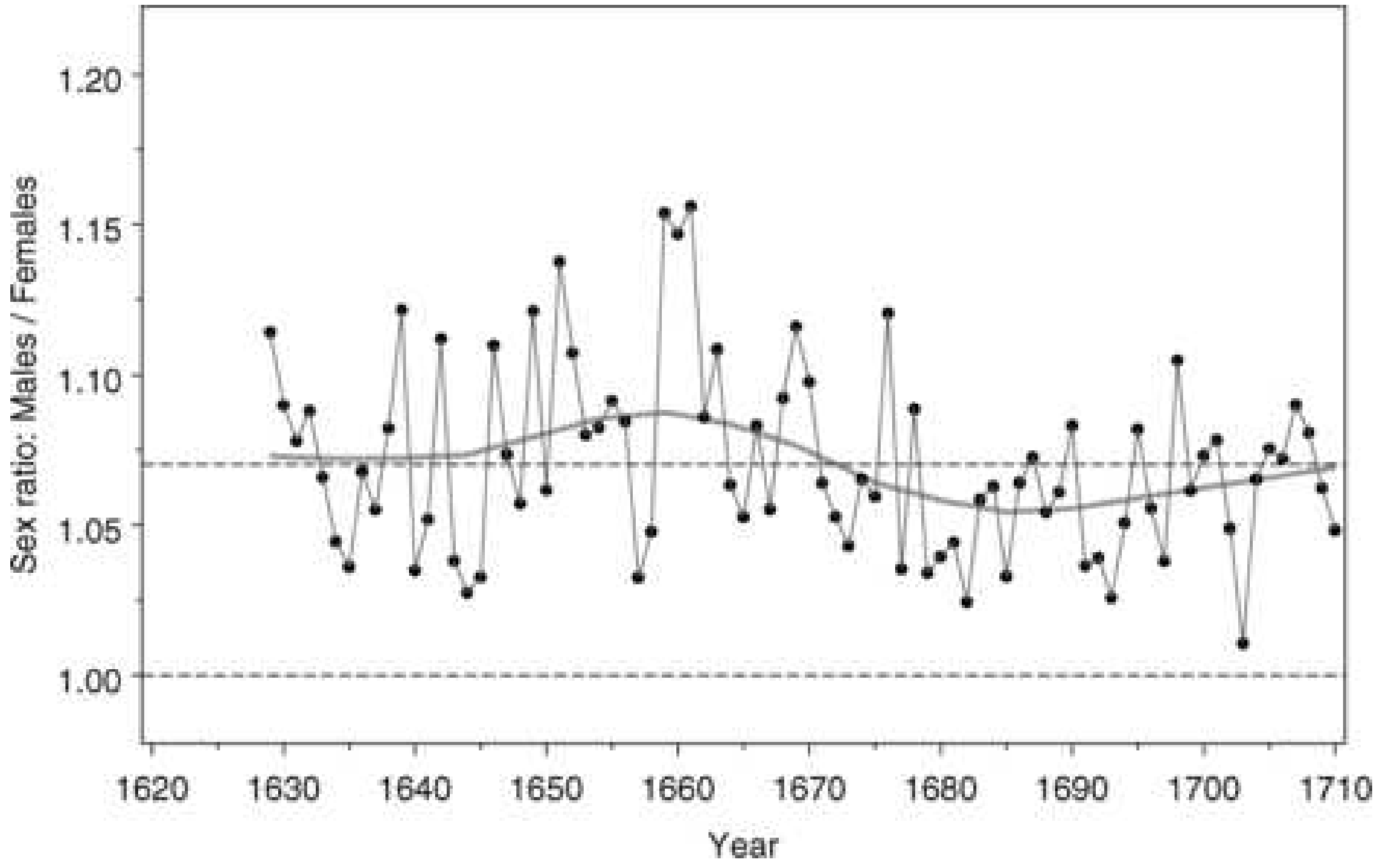

FIG. 1. *Arbuthnot's data on the male/female sex ratio. The average, shown by the upper dashed line is 1.07; the thicker line shows a loess smooth. The probability of ratios greater than 1 over 82 years is* $(\frac{1}{2})^{82} = 2 \times 10^{-25}$ *under the null hypothesis that the probability of a male birth is* 1/2.

system of crime reporting, collected quarterly from every département and recording the details of every criminal charge laid before the French courts: age, sex and occupation of the accused, the nature of the charge and the outcome in court.[1] Annual statistical publications of this data, known as the *Compte Général de l'Administration de la Justice Criminelle en France*, began in 1827 under the initiative of Jacques Guerry de Champneuf, the director of *affaires criminelles* in the Ministry of Justice (Faure, 1918, pages 293–294).[2]

During the same period, a wealth of other data on moral and other variables became available: data on age distributions and immigrants in Paris began with the 1817 census; Alexander Parent-Duchatelet (1836) provided comprehensive data on prostitutes in Paris, by year and place of birth; the Ministry of War began to record data on conscripts who could read and write; information on wealth (indicated by taxes), industry (indicated by patents filed) and even wagers on royal lotteries became available for the départements of France in various bulletins of the Ministry of Finance, 1820–1830. Thus, the stage was set by this "avalanche of numbers" (Hacking (1990)) for someone to try to make sense of competing claims about the causes of crime from comprehensive data and detailed analysis. André-Michel Guerry happened to be in the right place at the right time, but more importantly, he had a passion for numbers and his quest for taming it would nearly occupy his entire professional life.

[1]This extended a model begun in Paris in 1821 with annual publications of the *Recherches Statistiques sur la Ville de Paris et le Département de la Seine* under the direction of Jean Baptiste Joseph Fourier (1768–1830) toward the end of his life. These volumes detailed births, marriages and deaths, but also provided extensive tabulations and breakdowns of inmates of Parisian insane asylums, and of motives and causes of suicides. See Hacking (1990, pages 73–77).

[2]Guerry de Champneuf was said to be related to André-Michel, by a contemporary reviewer of the *Essai* (Caunter (1833)); Hacking (1990, page 77) calls him a cousin. There is no evidence to support a family relation. The similarity of names caused some confusion among American sociologists, starting with M. C. Elmer (1933), who conjoined the two into the siamese twin, M. de Guerry de Champneuf, or M. de Champneuf for short.

## 2. GUERRY'S LIFE, WORK AND METHODS

Unlike Quetelet, a brilliant academic politician and effective self-promoter who achieved prominence in academic and social circles throughout Europe and whose life has been widely biographed, Guerry's fame in his lifetime, like his life itself, was more modest. Aside from brief, bare-bones entries in the *Grand Dictionnaire Universel* (Larousse (1866)) and similar sources (Carré de Busserolle (1880); Vapereau (1858)), the primary sources on Guerry's life are the seven-page necrology by Alfred Maury, a long-time friend and fellow member of the Académie des Sciences Morales et Politiques, read at his funeral in April, 1866, and notices on Guerry's work by Hypolyte Diard and Ernest Vinet. These were initially published separately in the month of Guerry's death and then printed together in Diard (1867). Secondary sources include Whitt's (2002) preface to the translation of the *Essai*, Beirne (1993b) and a scattering of brief mentions, often in relation to Quetelet, by criminologists, sociologists (Lazarsfeld (1961); Isambert (1969)) and historians (Hacking (1990); Porter (1986)).

Guerry was born in Tours on December 24, 1802; his birth certificate lists his father, Michel Guerry, as a public works contractor, and the indications from historical sources are that his family circumstances were comfortable though neither wealthy nor highly connected. He studied law, literature and physiology at the University of Poitiers, and was admitted to the bar in Paris, where he became a Royal Advocat (Diard (1867)). In 1827 he began to work with the *Compte Général* in the course of his duties with the Ministry of Justice. He became so fascinated by these data that he abandoned active practice in law to devote himself to their analysis, a task he would pursue until his death in 1866. Sadly, no details of his personal or family life are available.[3]

One early statistical work (Guerry (1829)) examined the relation between weather and mortality from various diseases, and contained graphs of admissions to hospitals and polar area diagrams of the variation of weather phenomena, by month and hour[4]; other early studies concerned physiological characteristics (e.g., pulse rate) of inmates at insane asylums and prisons. Toward the end of his career, he invented a calculating or tabulating device (the *ordonnateur statistique*) to aid the work on the data from his last work and magnum opus (Guerry (1864)), the details of which remain shrouded.[5] Over his career, he produced the three major works on moral statistics described below. General discussion of his methods of analysis follows in Section 2.4.

### 2.1 1829: Statistique Comparée de l'État de l'Instruction et du Nombre des Crimes

Guerry's first publication on moral statistics was a large, single-sheet set of three shaded maps comparing crime and instruction titled *Statistique Comparée de l'État de l'Instruction et du Nombre des Crimes* produced together with the Venetian geographer Adriano Balbi (Balbi and Guerry (1829)), shown here in Figure 2. The data on crime from the *Compte Général* of 1825–1827 were combined with data from the census to give measures of population per crime (number of inhabitants to give one condemned person) for 81 départements; the data on instruction are based on the number of male children in primary schools in 26 educational districts (cours royales and académies) in France, also in the form of inhabitants per student.

[3]Guerry had no siblings, he never married and had no children. Nothing is yet known about his personal life in Paris. However, his family had deep roots in the area around Tours that have now been traced back to the early 1600s (Friendly (2007c)).

[4]These 1829 polar area charts predate those by Florence Nightingale (1857), who is widely credited as the inventor of this graphical form, using sectors of fixed angle, but varying radii to show frequency of some events, typically for cyclic phenomena. Guerry's plate shows six such diagrams, all for daily phenomena. Four of these show direction of the wind in 8 sectors, for 3-month periods; two show births and deaths, respectively, by hour of the day. Guerry says that these represent just a part of his original, much larger diagrams, which were not at first designed to be published.

[5]The *ordonnateur statistique* is simply mentioned in passing by Larousse; Maury (page 5) says the machine was offered by Guerry's heirs to the Conservatoire des Arts et Métiers. It is possible that this device briefly came into the hands of Maurice d'Ocagne, Professor at the ENPC and principal developer of nomography, but almost certain that he observed it when he discovered the collection of calculating machines held by the Conservatoire. Ocagne presented several lectures at the Conservatoire in February and March 1893 titled "Le Calcul Simplifié par les Procédés Mécaniques et Graphiques," but there is no mention of Guerry in the papers printed in the *Annales du Conservatoire* (d'Ocagne (1893)). There is a curious connection with IBM here: IBM France introduced the term *ordinateur* to replace the deprecated franglais term, *computeur*; Ocagne also studied another collection of early calculating devices belonging to General Sebert, purchased later by IBM France, which may also have acquired others. The archivists at IBM can find no records of these.

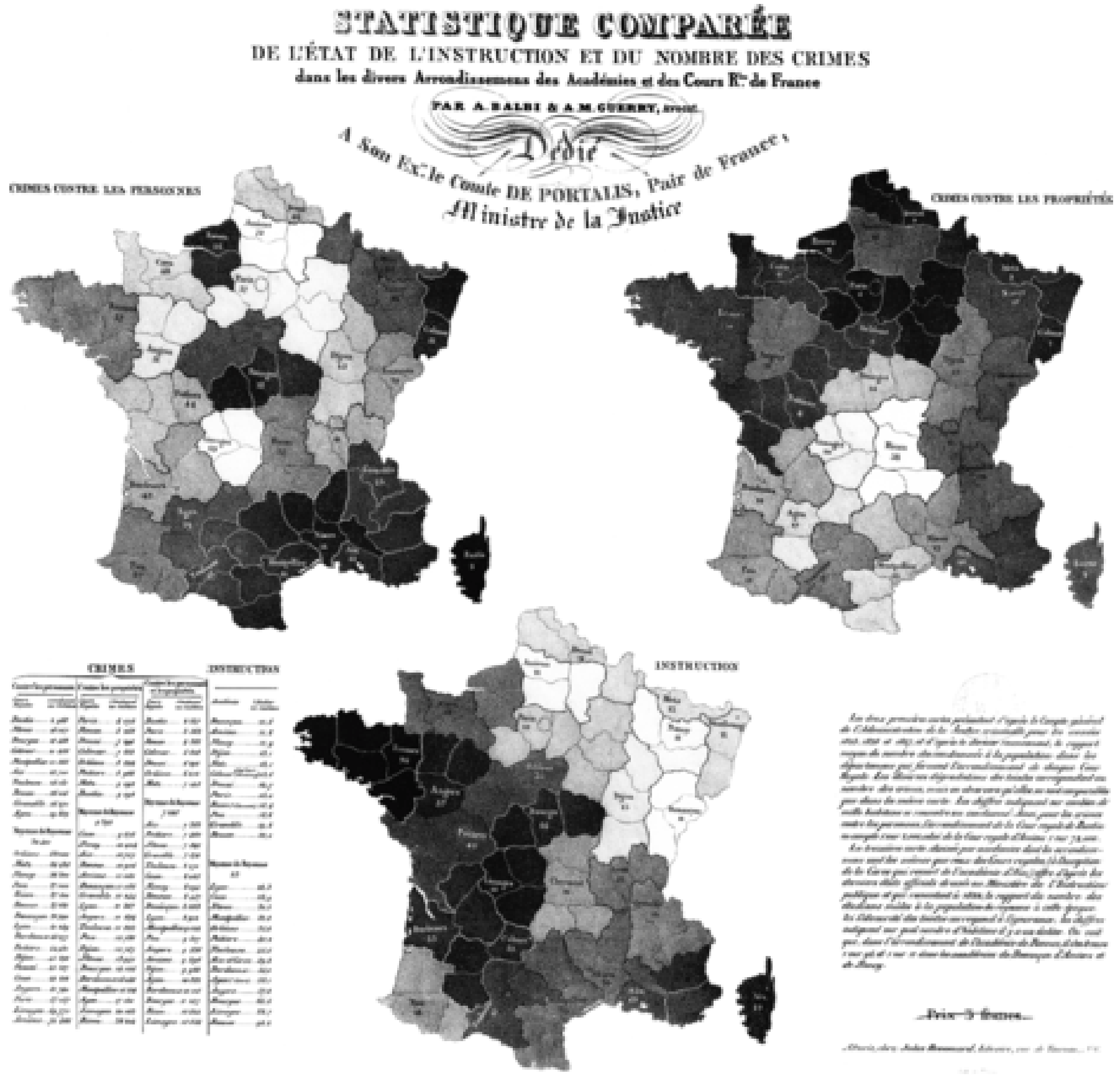

FIG. 2. *Guerry and Balbi's 1829* Statistique Comparée de l'État de l'Instruction et du Nombre des Crimes. *Top left: crimes against persons; top right: crimes against property; bottom: instruction. In each map, the départements are shaded so that darker is worse (more crime or less education). The legend at the lower left gives the data on which the maps were based.* Source: *Courtesy of BNF; Palsky (1996, Figure 19).*

The exact source of the data on instruction is unclear. The legend for the maps cites the Ministry of Instruction for 1822, but Guerry's printed commentary on this work (republished in Guerry (1832)) cites Balbi's 1822 *Statistique du Royaume de Portugal...* as the first document that published a table of data on the level of public education in France.

What is clear is the impact they had on Guerry and others. In 1823 the geographer Conrad Malte-Brun (1775–1826)[6] in commentary on Balbi's table remarked that there appeared to be a much lower level of instruction in the south of France compared with the north; he referred to this as the contrast between *la France obscure* and *la France éclairée*. Quite shortly, Baron Charles Dupin, perhaps inspired by this observation, thought to portray these data on a map of France (Dupin (1826), 1827), using shades of varying darkness to depict degrees of ignorance. This invention—the first modern statistical map—was the starting point of a true graphical revolution that Guerry would extend to a more general social cartography with the *comparative* analysis of social issues. The legacy of this revolution is commonplace today, in maps of disease incidence, poverty, child mortality, income distribution[7] and so forth (Friendly and Palsky (2007)).

[6] *Journal des Débats*, 21 jul 1823, pages 3–4.

Guerry and Balbi's *Statistique Comparée...* in 1829 was the first use of shaded maps to portray crime rates. Their presentation is also notable in the history of statistical graphics as the first to combine several moral variables in a single view, allowing direct comparison of crimes against persons and against property with data on instruction across the départements of France. They suggested that, (a) surprisingly, personal crimes and property crimes seemed inversely related overall, but both tended to be high in more urban areas; (b) a clear demarcation could be seen in instruction between the north and south of France along a line running northeast from Geneva (Ain) to Côtes du Nord[8]; (c) the north of France not only showed the highest levels of education, but also of property crime. At the very least, this work testified to the importance of detailed data, sensibly presented, to inform the debate on the relations of crime and education.

### 2.2 1833: Essai sur la Statistique Morale de la France

Over the next three years Guerry would occupy himself with the extension and refinement of these initial results, with extensive tabulation of new data from diverse sources and with answers to methodological questions that might lead to challenges to his conclusions.

On the methodological side, he discussed how these measures should be defined to ensure comparability across France and what we would now term validity of the indicators used. For education, for example, he considered the reported levels of instruction (number of male children in primary school) to be suspect due to variations in local reporting; after 1827, better and more uniform data became available from the Ministry of War, whose exams for new recruits gave numbers for those who could read and write.

A second major question he addressed was whether crimes should be counted by the number of indictments (*accusés*) or by the number of convictions (*condamnés*). Here, he argued persuasively that the number of indictments was a more useful indicator of the number of crimes committed because it is less likely to be influenced by the factors that affect juries: the nature of the crime, severity of punishment and the place where the accused is judged. Moreover, although an indictment by no means implies the guilt of the accused, it does reasonably imply that a crime was committed; conversely, a person might be acquitted for a variety of reasons, but that does not mean that a crime did not take place.

The *Essai* published in 1833 contained numerous tables giving breakdowns of crimes against persons and property by characteristics of the accused, frequencies of various subtypes of crime in rank order for both men and women (for men, the most popular personal crime was assault and battery; for women, infanticide) and frequencies of crimes by age groups.

To go beyond simple description, Guerry classified the crimes of poisoning, manslaughter, murder and arson according to the apparent motive indicated in court records (for poisoning, the motive was most frequently adultery; for murder, it was hatred or vengeance). This quest to examine motives and causes is most apparent and impressive in his analysis of suicide, a topic of considerable debate in both the medical community (which considered it in relation to madness and other maladies) and the legal community (which considered whether it should be a crime or at least within the purview of the justice ministry). "What would be useful to know would be the frequency and importance of each of these causes relative to all the others. Beyond this, it would be necessary to determine whether their influence . . . varies by age, sex, education, wealth, or social position" ((Guerry, 1833, page 131, WR trans.)). To this end, he carried out perhaps the first content analysis in social science by classifying the suicide notes in Paris according to motives or sentiments expressed for taking one's life. This approach to the study of suicide would later be adopted by Durkheim (1897), but without much credit to Guerry and other moral statisticians.

The *Essai* also contained a collection of bar graphs, highlighting certain comparisons (crimes against persons occurred most often in summer months, while those against property were most frequent in the winter; suicides by young males were most often carried out with a pistol, while older males preferred hanging). As well, to discuss geographical differences and relate these moral variables to each other, he prepared six thematic maps of France, adding illegitimate births (*infants naturelles*), donations to the poor (number of gifts and bequests) and suicide to those of personal crime, property crime and education presented earlier, but based on more complete data and better indicators. For ease of comparison, these variables were all expressed in a form such that "more is better," for example, population per crime or percent able to read and write. In preparing the maps, these variables were first converted to ranks and then the départements were shaded according to rank, so that darker tints were applied to the départements that fared worse on a given measure (more crime, less education).

[7]See www.worldmapper.org for a collection of over 300 world cartograms, where territories are resized according to the subject of interest.

[8]This sharp cleavage between France du Nord et Midi or France obscure vs. France éclairée would become reifined as the "Saint-Malo–Geneva line" and generate much debate about causes and circumstances through the end of the 19th century.

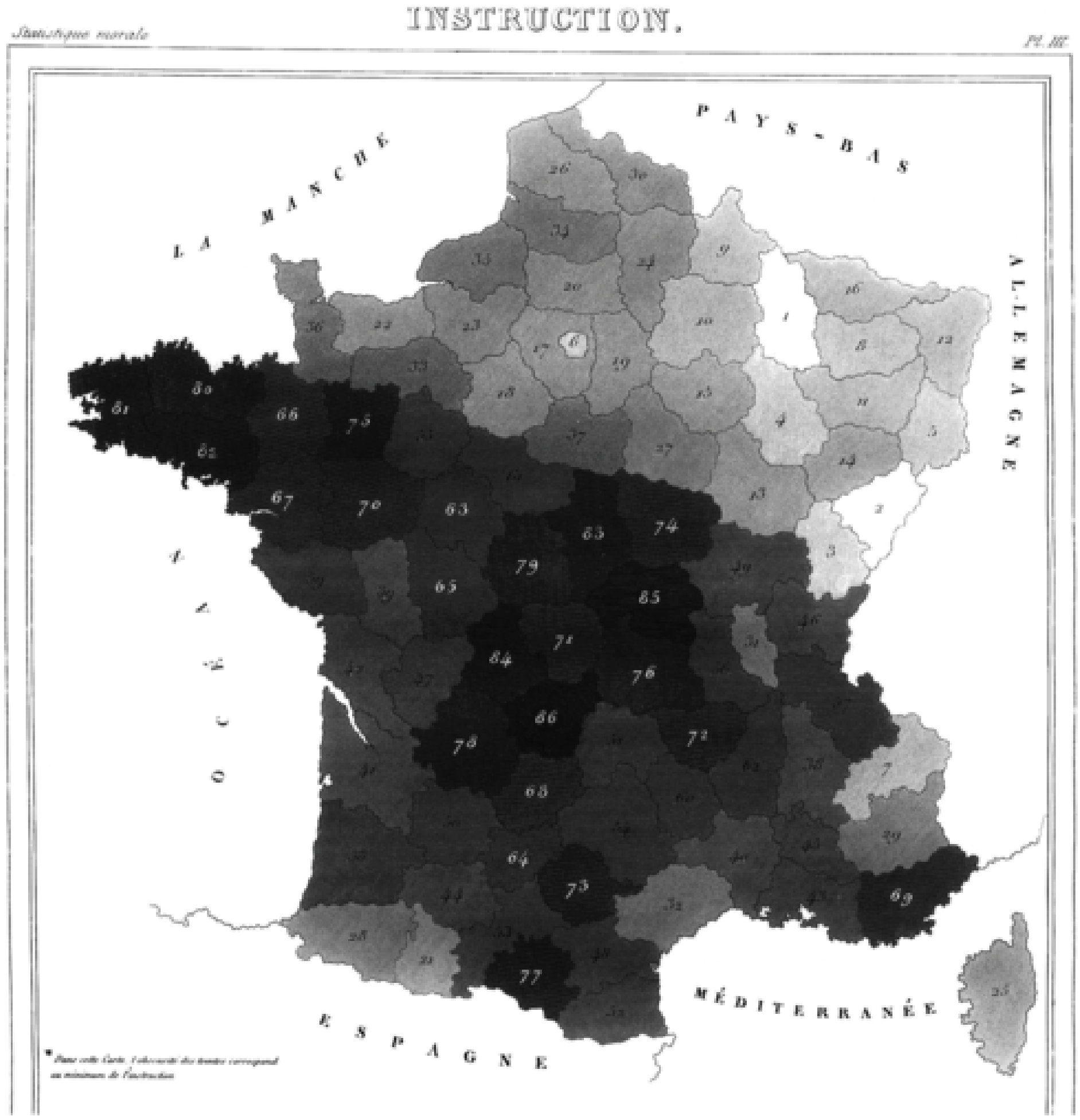

FIG. 3. *Guerry's 1833 map of levels of instruction in France (Plate* III*). The original contains the numbers (percent of young men who can read and write) for each département in rank order listed below the map.* Source*: Author's image collection.*

These maps are generally more detailed and finely drawn than those produced in 1829. Figure 3 shows an example, the map labeled "Instruction," but showing the percent of military conscripts who could read and write. Figure 4 shows all six of Guerry's maps, reproduced from his data using modern software.

Guerry's *Essai* was received with considerable enthusiasm in European statistical circles, particularly in France and England. In France it was awarded the Prix Montyon in 1833 and the publication includes a laudatory report on its contents to the Académie des Sciences. Guerry was also elected to the Académie des Sciences Morales et Politiques and at some point was awarded the cross of chevalier of the Legion of Honor (Diard (1867)). The *Essai* was reviewed quite favorably in the *Athenaeum* (Caunter (1833)) and the *Westminster Review* (Anonymous (1833)), which gave a lengthy discussion of Guerry's findings and praised the book as one of "substantial interest and importance." Henry Lytton Bulwer's (1834) *France*, *Social*, *Literary*, *Political* devoted 26 pages to Guerry's results, calling the work "remarkable on many accounts." Guerry displayed the maps in several expositions in Europe and, in 1851, had two exhibitions in England—an honored public one in the Crystal Palace at the London Exposition and a second at the British Association for the Advancement of Science (BAAS) in Bath.

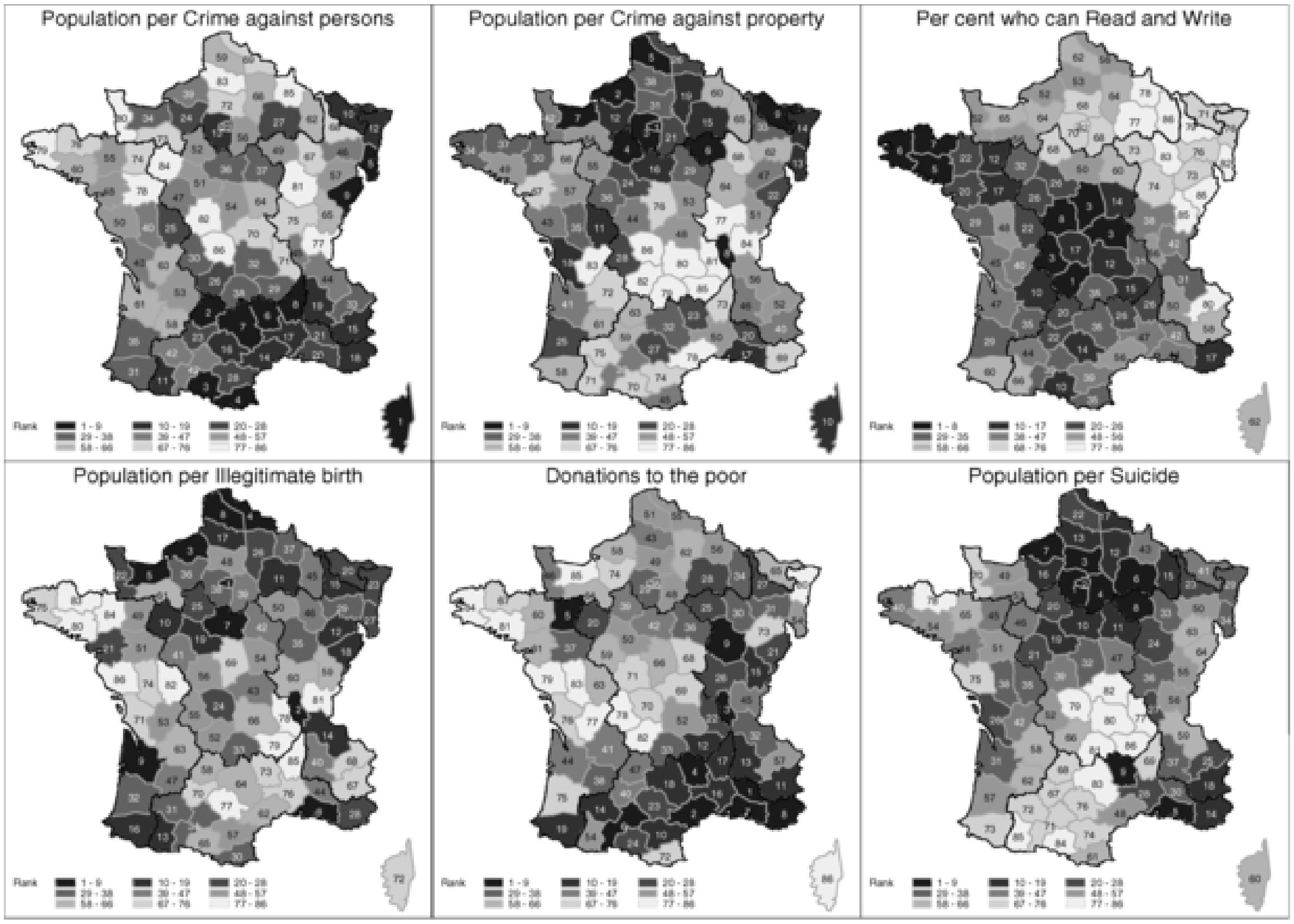

Fig. 4. *Reproduction of Guerry's six maps. Color coding, as in Guerry's originals, is such that darker shading signifies worse on each moral variable. Numbers for each département give the rank order on that variable.*

### 2.3 1864: Statistique Morale de l'Angleterre Comparée avec la Statistique Morale de la France. . .

Guerry's most ambitious work, and the capstone of his career, did not appear for another 30 years, but it was well worth waiting for. The *Statistique Morale de l'Angleterre Comparée avec la Statistique Morale de la France* was published in a grand format ($56 \times 39$ cm, about the size of a large coffee table); it contains an introduction of 60 pages and 17 exquisite color plates. The introduction sets out Guerry's view of the history of the application of statistics to the moral sciences. In it, he proposes to replace the term "moral statistics" or simply documentary statistics with "analytical statistics." The former, presented almost invariably in tables, is concerned with the numerical exposition of facts; the latter presents the successive transformation of these facts, by calculation, by concentration and their reduction to a small number of general abstract results. One can see here a thorough explanation of the graphic method applied to moral and social data.

Fifteen of the plates show data for the départements of France (from 1825 to 1855) or the counties of England (1834–1856) on a particular topic, first for France, then for England: crimes against persons, crimes against property, murder, rape, larceny by servants (*vol domestique*), arson, instruction and suicide (only for France). In each case, to ensure comparability of the numbers for the various crimes across départements and counties, and from one measure to another, Guerry standardized the rates for each map and country to "degree of criminality," with a mean of 1000 and common (unspecified) metric. Thus, one could easily see where Paris/Seine or London stood on murder vs. suicide or compare one to the other on theft.

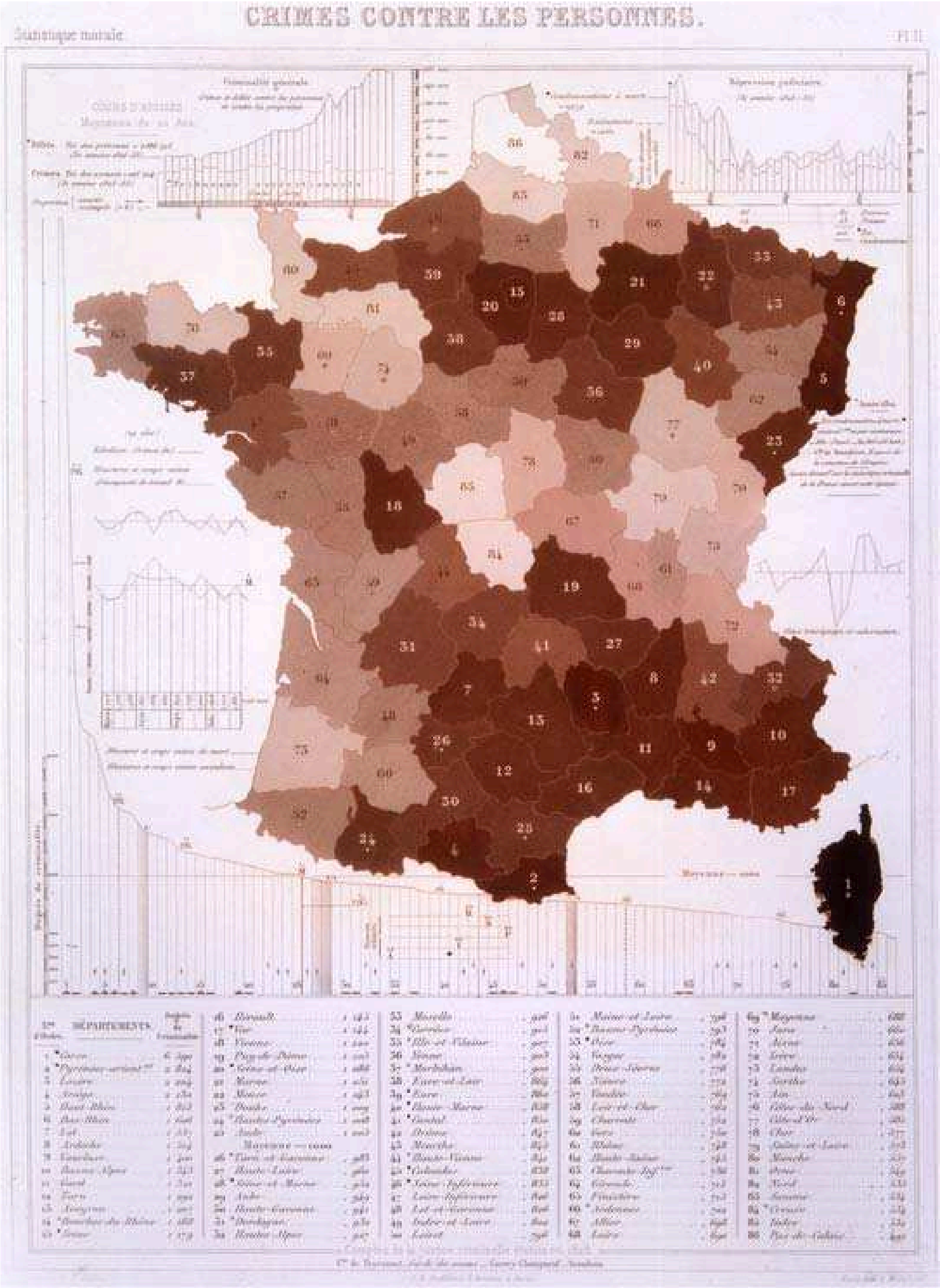

FIG. 5. *Guerry's 1864 Plate 1: Crimes against persons in France.* Source*: Courtesy of Staatsbibliothek zu Berlin.*

Each of these plates exemplifies the program of *statistique analytique* that Guerry had in mind, as illustrated by Figures 5 and 6. The map of England or France shows the geographic distribution, with counties or départements shaded according to their rank order on the variable, the highest (rank = 1) shaded darkest and the lowest shaded lightest. A large variety of special symbols and annotations are used on the map to indicate noteworthy patterns or circumstances, for example, up or down arrows to show increase or decrease over time in a geographic unit. The table below the map lists the ranks and data values, expressed as "degree of criminality."

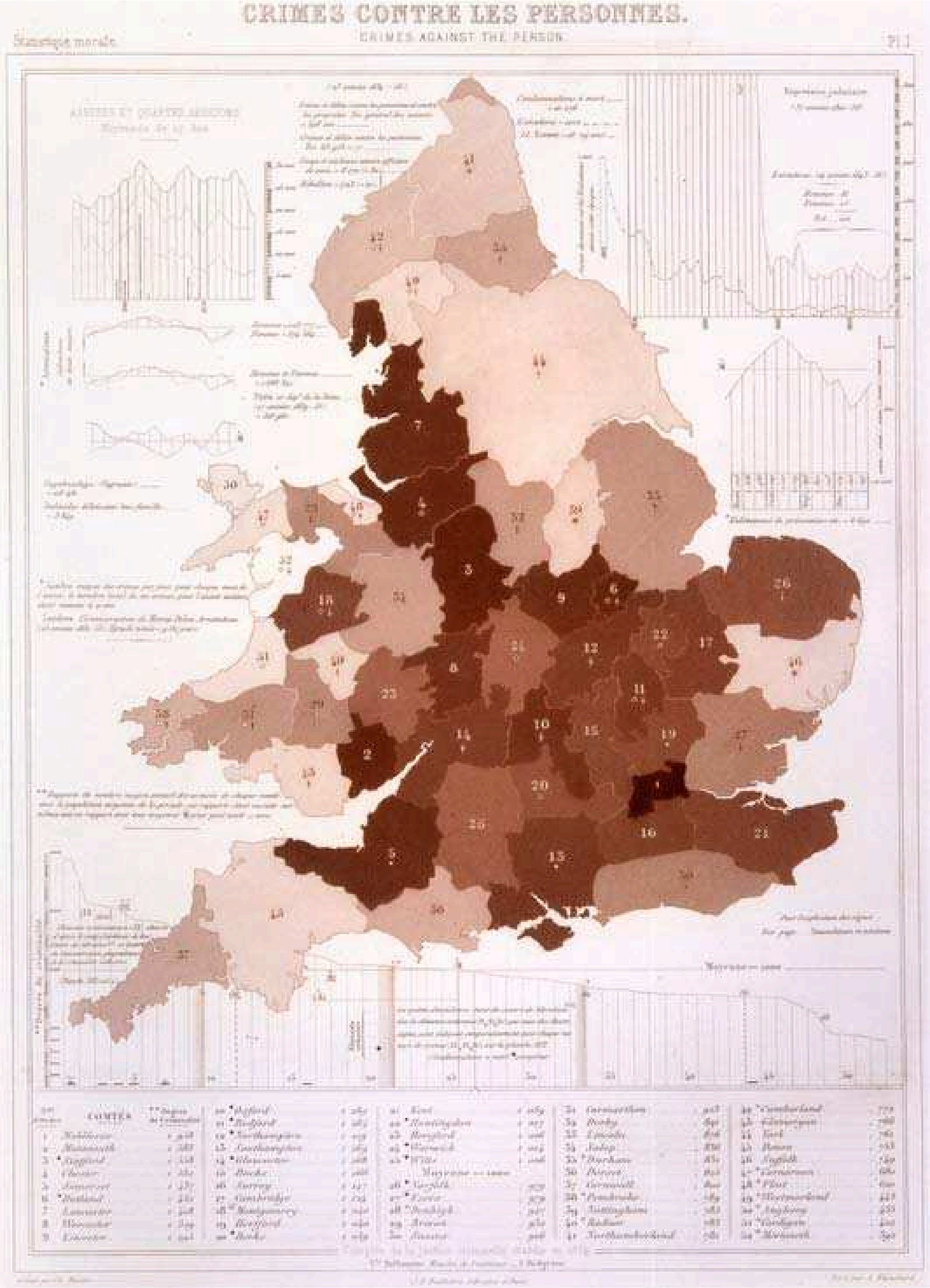

FIG. 6. *Guerry's 1864 Plate 2: Crimes against persons in England.* Source*: Courtesy of Staatsbibliothek zu Berlin.*

Each map is an overall summary for 30 years, for all accused and for all crimes in a given class. Guerry wanted also to show patterns, trends or deviations within these data. Thus, surrounding each map, he placed a variety of line graphs designed to decompose or transform these overall facts or to relate them to other factors. Most of these featured time series graphs of trends over time, often decomposed into separate series by subtype of crime or characteristics of the accused, such as age or sex. Plate 1 (Figure 5), for example, includes at the top a general summary time series of the number of crimes against persons and property (left), and of the numbers condemned to death and executed over the 30 year period (right: Figure 7). Distributions of crimes by month of the year, moreover, reveal that crimes against persons were greatest in the summer and least in the winter; property crimes in France showed the reverse pattern. Beneath the map, an index plot of the degree of criminality values by rank shows the form of the distribution across counties and départements. Again, many special symbols are used to mark the minimum, maximum, mean, median, increase or decrease, possibly fallible numbers and so forth; the nearly 100 such symbols defined in an appendix clearly required some typographic calisthenics, as they run through several alphabets plus the available diacritical marks.

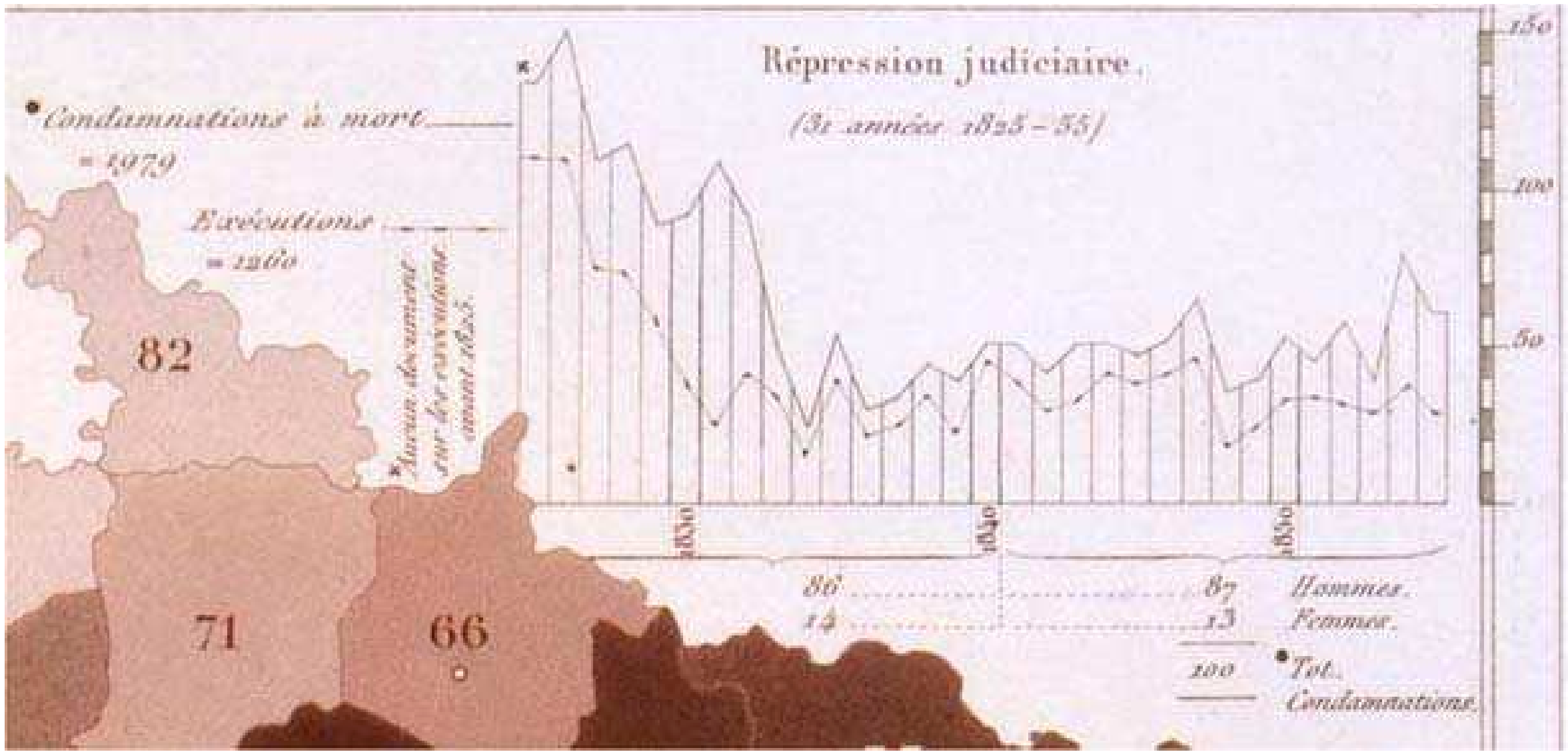

FIG. 7. *Detail from Plate 1: Time series chart of death sentences and executions, 1825–1855.* Source*: Courtesy of Staatsbibliothek zu Berlin.*

The final two plates serve as the culmination of Guerry's program of analytical statistics and provide an ambitious attempt to delineate multifactor and multivariate relations among rates of crime in England and France; these are discussed in the subsection below.

One cannot fail to be impressed by the sheer volume of data summarized here; these include over 226,000 cases of personal crime in two countries over 25 years and over 85,000 suicide records, classified by motive. Guerry estimated that if all his numbers were written down in a line, they would stretch over 1170 meters! Hacking (1990, page 80) credits this observation as the source of his phrase "an avalanche of numbers."

### 2.4 Guerry's Methods and Analyses

Guerry worked in a time before the ideas of correlation and regression were invented, and at about the same time that the first true scatterplot of two variables appeared in an astronomical paper by John Herschel (1833); see Friendly and Denis (2005) for an account of the origin of the scatterplot). Although he later included quotations from Herschel in the 1864 comparative study of England and France, there is no evidence that he was aware of any bivariate methods in 1833, and even the 1864 work shows no awareness of scatterplots for studying the relation between variables.

2.4.1 *Graphic comparisons.* His method was therefore limited to direct comparisons of distributions or pairs of variables, shown either in ranked lists or on his shaded maps. Figure 8 attempts to capture the spirit of Guerry's approach, comparing the map of rates of crimes against persons with that for literacy. To illustrate Guerry's thinking, I used the interactive Mondrian software (Theus (2002)) to compose side-by-side maps with a ranked parallel coordinates plot in the middle. The best conclusion one can draw directly from the maps is that any relation between the two is weak or inconsistent, as there are regions where literacy is low and crime is high (central France), but other areas where both are relatively high or low. He says:

> Let us now compare this map with the one for crimes against persons. The maximum crime rate is in Corsica. Is this because there is greater ignorance there? Our map supplies evidence to the contrary. Further, the minimum occurs in the western and central provinces. Can it be said that the highest level of education prevails there? Clearly, the relationship people talk about does not exist (Guerry (1833), page 90, WR trans.).

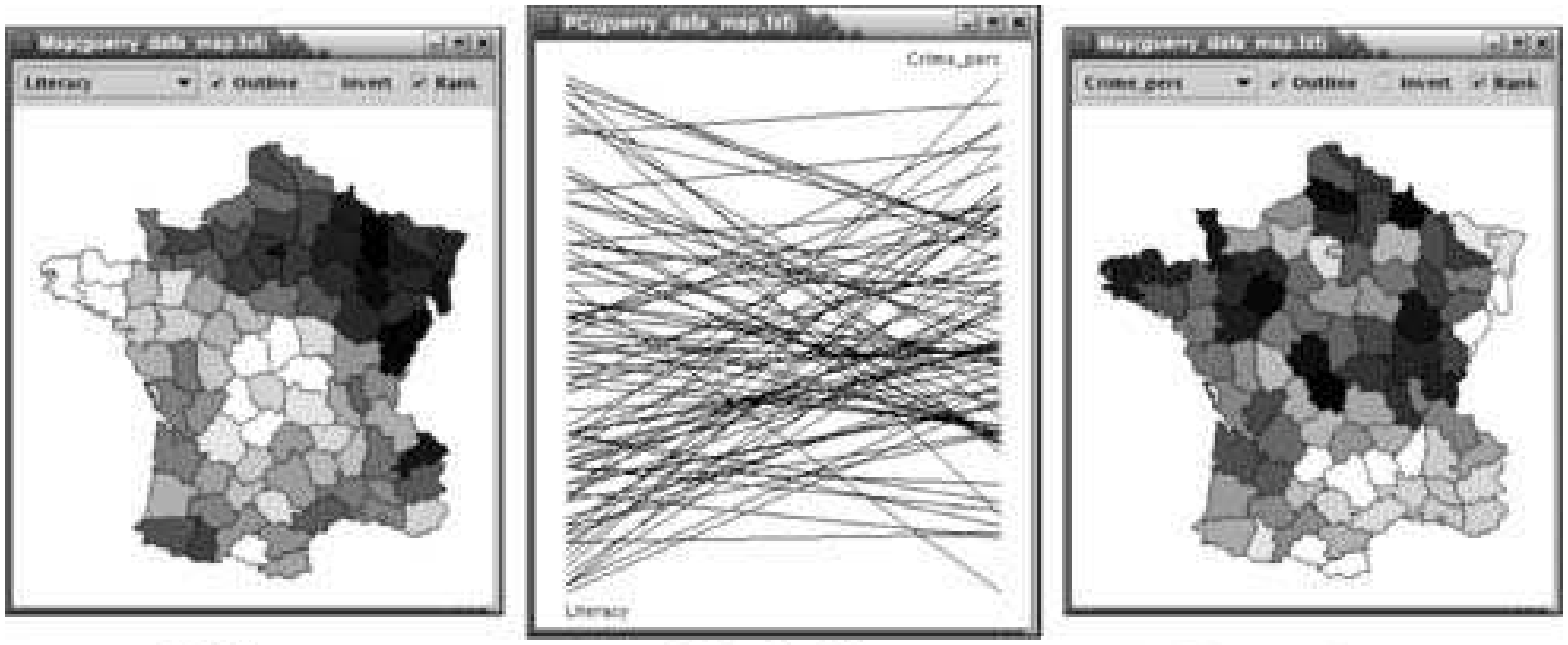

(a) Literacy (b) Ranked lists (c) Crimes against persons

FIG. 8. *Comparison of crimes against persons with literacy (% who can read and write). The two maps* (a) *and* (c) *show, by shading intensity, the rank orders of each of these variables. The middle panel* (b) *is a parallel coordinates plot of the ranks, connecting the two maps. Images from analysis with Mondrian (Theus (2002)).*

Similarly, in discussing the relationship between crime and suicide he says:

> One might think that ... the violent national character of our southern provinces, which ... produce such a large number of crimes against persons, should also lead many people to kill themselves. But this would be incorrect. A comparison of the suicide map with the one for crimes against persons leads to the discovery that, with very few exceptions (especially for Alsace and Provence), the départements where the lives of others are most often attacked are precisely those where people most rarely make attempts on their own and vice versa (Guerry (1833), page 130).

Today, conclusions based on such comparisons of grouped rates are often charged with commission of an ecological fallacy (Freedman (2001))—that inferences about relationships observed for aggregate data may not hold for individuals (due to confounding or aggregation bias).[9] Guerry was certainly aware of the problem of ecological correlation, at least in general terms. In his discussion of the relationship between crimes and education (Guerry (1833), pages 94–95), he noted that since 1828 the *Compte* presented data on the educational level of accused persons and asked rhetorically, "is it the case that there is indeed greater ignorance among individuals prosecuted for crimes against persons than among other defendants?"

One answer to the problem of ecological correlation is provided by comparing the literacy of prisoners with those in the general population. To counter the argument that those found guilty of crimes against persons are more ignorant than those who commit only property crimes, he showed that educa-tion is in fact *higher* among those committing personal crimes; moreover, "among these latter crimes, the most depraved and perverse appear generally to be committed by preference by educated perpetrators" (Guerry (1833), page 94).

[9]The classic example is Durkheim's (1897) assertion that suicide was somehow promoted by the social conditions of Protestantism because suicide was higher in countries that were more heavily Protestant. Of course, the largely Protestant countries differed from the Catholic countries in many ways besides religion. Current research on the relation of suicide to other variables continues to examine correlations over geographical units, often uncritically (e.g., Bills and Li (2005)).

To examine how the particular types of crimes committed varied with age of the accused, Guerry in 1833 prepared the ranked lists shown in Figure 9 for both crimes against persons and crimes against property. To make the trends more amenable to visual inspection, he connected the same crime across age with lines. This gives a semigraphic display that combines a table (showing actual numbers) with the first known instance of a parallel coordinate plot. In the original, the trace lines are hand colored in different light hues to make them visually distinct.

From this, Guerry discussed a variety of trends, such as the decrease in indecent assault on adults (*viol sur des adultes*) with age (while indecent assault on children rises to the top for those over 70) and the increase in parricide with age (surprisingly reaching a maximum for "children" aged 60–70). Among crimes against property, he pointed out that theft and domestic theft are the most common at all ages, but (the curious subcategory) theft from churches has a U-shaped relation with age, while extortion and embezzlement (*concussion*) rise from the very bottom in the young to near the top among older groups.

As noted above, by 1864, Guerry was striving for more analytic methods to reveal the regularity and variation in moral statistics, and he had an enormous amount of data: 30 years for France, 23 for England. In the final two plates in 1864, Guerry abandoned the geographical framework of the map to highlight more general patterns in crime and relations with explanatory and possibly causal factors, and how these compare in England and France.

Plate 16 (shown in Figure 10) is devoted to a detailed comparative analysis of the age distribution for various crimes and suicide (only for France, bottom left). In contrast to the ranked lists he had used earlier (Figure 9), he used side-by-side displays for England (197,000 accused of known age) and France (205,000 accused) of 10 collections of frequency distributions across age for crimes in various categories (theft, arson, murder, indecency and so forth). Each block provides separate curves comparing the age distributions of subtypes within a given category (e.g., Block II: murder, manslaughter, grave wounding; Block VI: burglary, housebreaking, robbery). Various annotations on the charts show the mean (M), missing data (small circles, indicating no crimes recorded in a given age category), the relative frequency of crimes committed by those under 21 and so forth.

2.4.2 *Guerry's Magnum Opus.* The last plate (Plate 17, shown in Figure 11), titled *Causes Générales des Crimes*, is by far the most impressive and also the most complex, and may be called Guerry's Magnum Opus. It is a novel semigraphic table devised to show the multivariate associations of various types of crimes with other moral and population characteristics, but at a time when even bivariate methods were unknown. The image shown here cannot do justice to the original, so I will try to describe it and also convey the sense of awe I felt when I first saw it in the British Library. Guerry's goal here is to show the factors associated both positively and negatively with crimes and their geographic distribution, using data from England as a specimen of this approach.

The chart is headed "Libration Comparée des Crimes de Chaque Nature et des Éléments Statistiques avec Lesquels ils sont Liés dans leur Distribution Géographique" (comparative libration of the crimes of each nature and the statistical elements with which they are linked in their geographical distribution). The rows show 23 types of crimes ordered by frequency and seriousness from top (debauchery, bigamy, domestic theft) to the bottom (fraud, rape, murder). Just below this are sets of summaries condensing these into crimes against persons and property, and other general categories. The columns refer to the rank orders of the 52 counties of England on each crime separately. Thus, on crimes against persons, Middlesex stands at rank 1 (left), with a "degree of criminality" of 1958, while Merioneth is at rank 52 (right) with 392. For the crime of arson, different counties occupy these ranks, of course, but it is the characteristics of the counties at various ranks that Guerry wants to show.

The entries in this graphic table are symbols for a variety of moral and social characteristics found either with high prevalence or low prevalence in the particular county at each rank for each crime. The legend at the bottom identifies the following kinds of symbols: (a) those for aspects of population (density of population, percentage of Irish, agricultural, maritime, domestics, etc.); (b) aspects of criminality (predominance of male, female, young, old, etc.

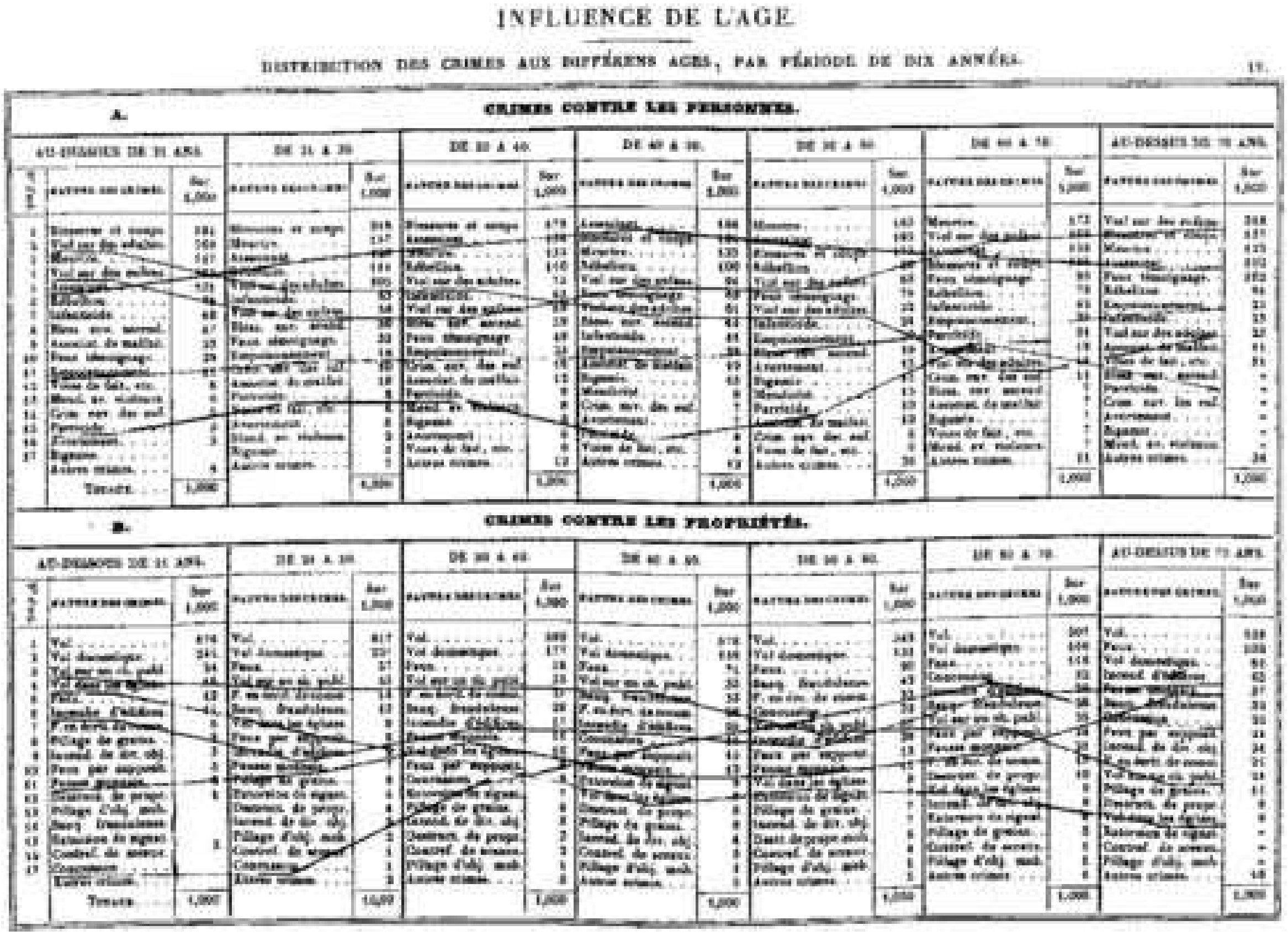

FIG. 9. *Relative ranking of crimes at different ages.*

relative to the average); (c) instruction (predominance of instruction of males, of criminals, of prisoners); (d) aspects of religion (Anglicans, dissidents, Catholic, etc.; attendance at public worship).

Overlaid on this are several sets of lines tracing profiles of the "centre de libration" (an astronomical term meaning center of oscillation) of various types of social indicator symbols. One curve for the symbol **a** (population density) is drawn as an example and labeled "path of **a** in the vertical series of ordinates," the idea being that one could see directly to which crimes population density was related positively (bigamy and domestic violence) and negatively (arson and cattle theft). Starting at the left (right) are two other smoothed curves labeled "curve of positive (negative) coincidence."

An inset quotation from J. W. Herschel (1831) in a box at the top right sums up Guerry's anticipation of the utility of his method, and a caveat: "Causes will very frequently become obvious by a mere arrangement of our facts in the order of intensity, though not of necessity, because counteracting or modifying causes may be at the same time in action."

It should be noted that Guerry intended this only as a specimen. He did not provide an analysis of these data or the obvious parallel chart for France, nor did he draw conclusions about the many relations between crimes and these social and moral aspects. He states in the introduction that such discussion will be the subject of a subsequent book, but this was never published.

This magnificent volume, published in 1864, had been crowned by the Académie in 1860 and was awarded the Prix Montyon the following year (Bienaymé (1861)). In October of 1864, Guerry, who had been made an honorary member of the Statistical Society of London (SSL), travelled to England to attend the BAAS meetings in Bath at the invitation of William Farr; president of the society, Farr had also been instrumental in arranging Guerry's access to the judicial records of England. The *Statistique Morale de l'Angleterre* and its splendid 17 plates were put on public display for the nearly 2800 members who attended, and became the subject of a commentary by W. Heywood, vice-president of the SSL.

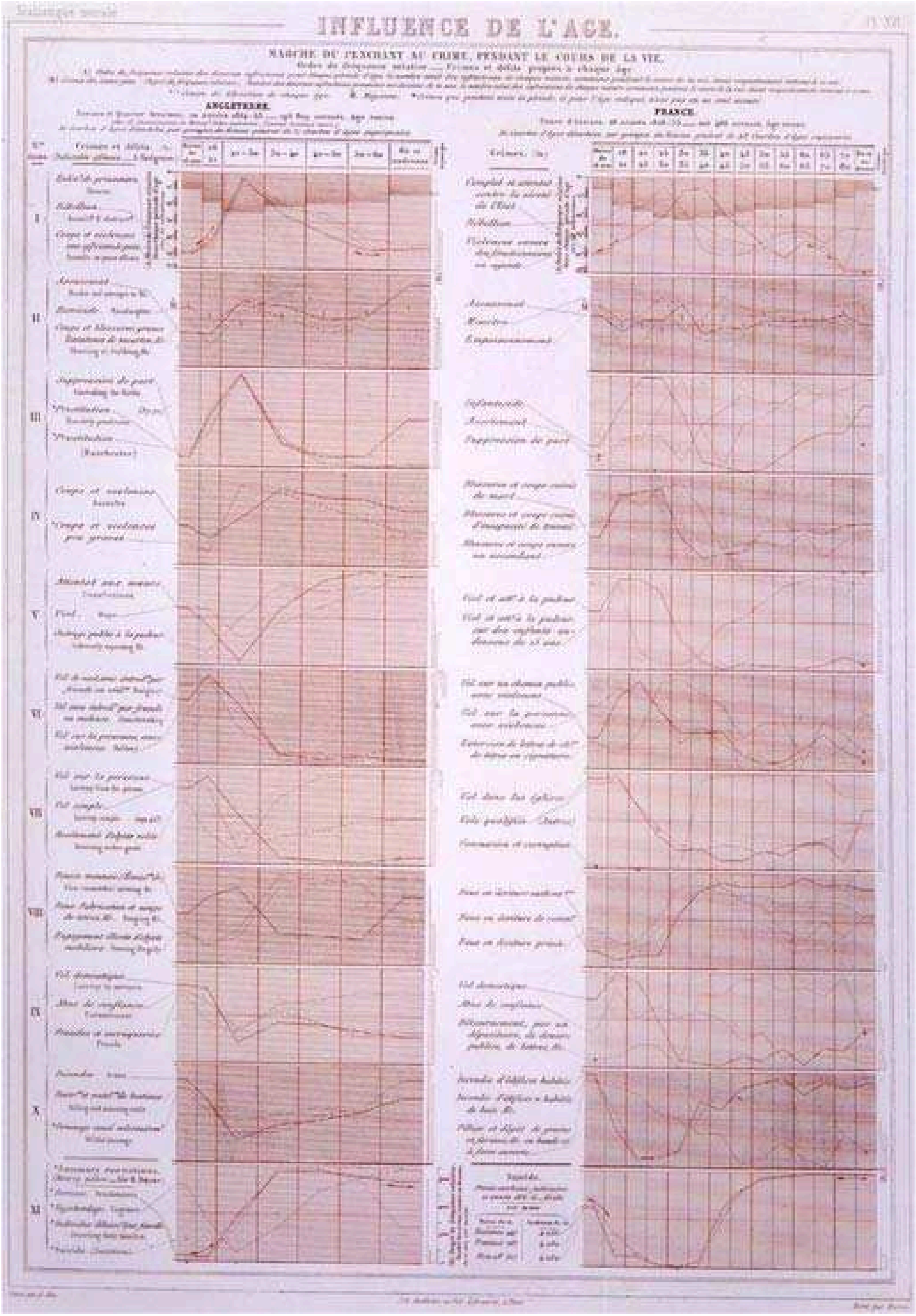

FIG. 10. *Guerry's 1864 Plate 16: Influence of age. Left: Age distributions for various crimes in England; right: in France.* Source*: Courtesy of Staatsbibliothek zu Berlin.*

The following August, while consulting the archives of the Hôtel de Ville de Paris, Guerry suffered a stroke; he survived, but grew progressively weaker, and died on April 9, 1866 at age 64. Some work on the ambitious analysis and interpretation of the data for England and France was continued and reported by Diard (1866), who described Guerry's last work as an atlas of criminal justice comparable in scope and beauty to the atlas of Berghaus (1838) on physical geography of the world. Guerry's papers were donated by his heirs, Charles and André Poisson, to the Société des Sciences, Arts et Belles-Lettres

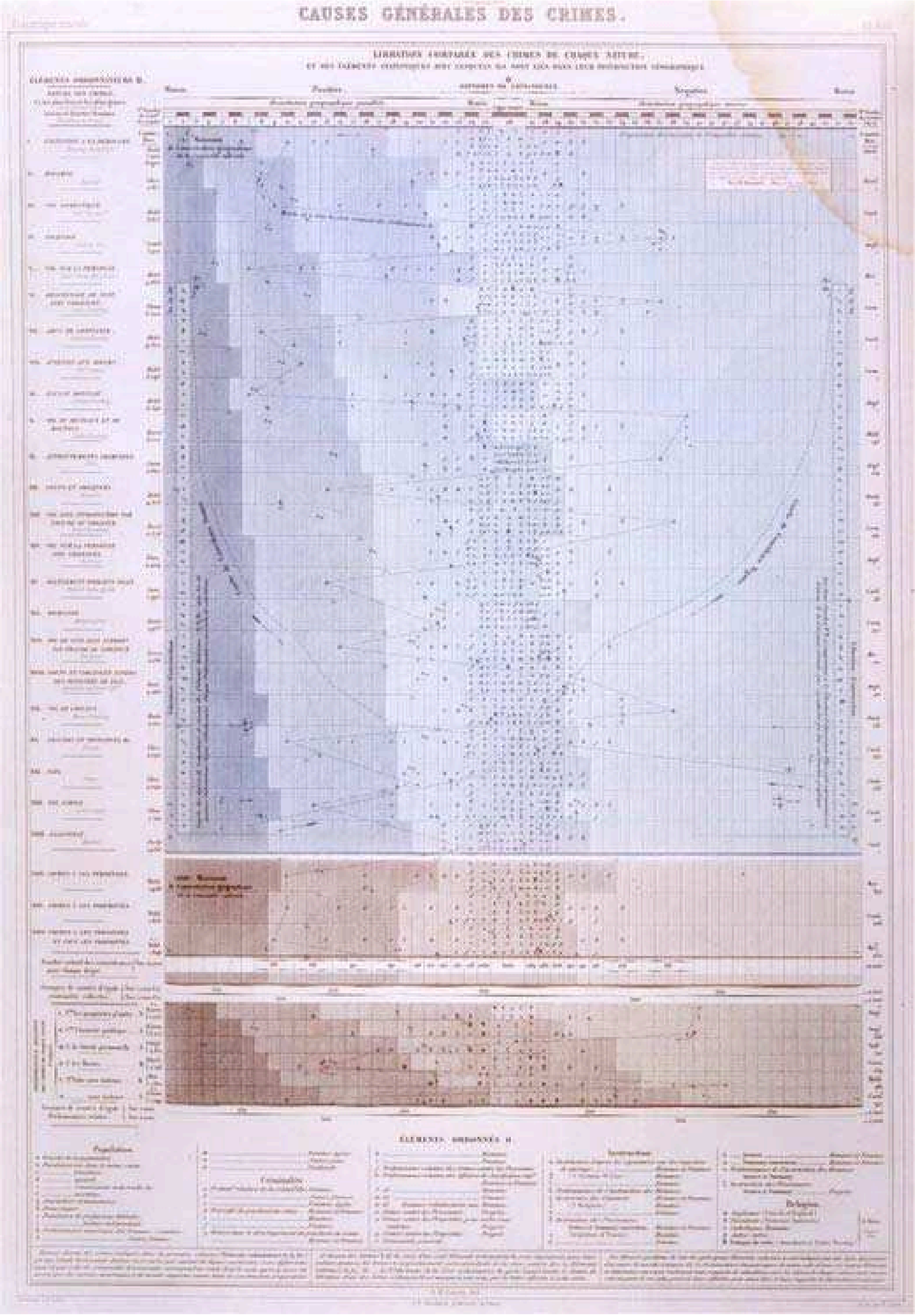

FIG. 11. *Guerry's 1864 Plate 17: General causes of crimes.* Source*: Courtesy of Staatsbibliothek zu Berlin.*

d'Indre-et-Loire; they have not been located, if they yet survive.[10]

[10]In subsequent work (Friendly (2007c)) I have determined that Guerry's notes and papers concerning the 1884 volume were entrusted to Diard. The *Annales de la Société...* record communications by Diard (April, 1867; 1868) concerning a project to edit and publish some of these manuscripts, adopted by the Société in 1869. This was never completed and no trace of these works has been found in local archives.

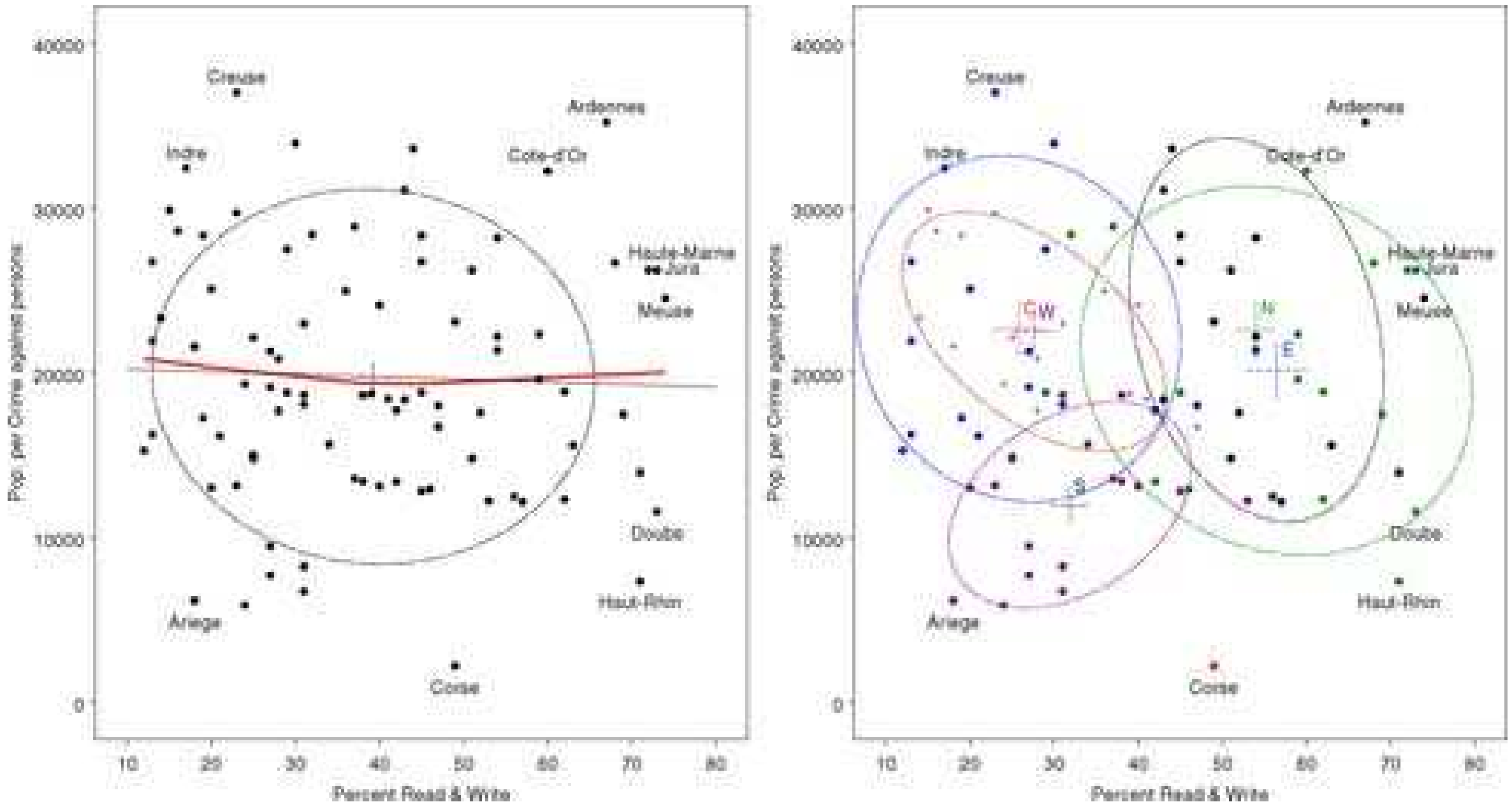

FIG. 12. *Plots of crimes against persons vs. literacy. Left: All regions, with 68% data ellipse and loess smooth. Right: Region differences. Départments outside the 90% data ellipse are identified in both panels.*

It is intentionally hyperbolic for me to call the *Essai* the book that launched modern social science, and to tell the story of Guerry's contributions to statistics and graphics, I have intentionally avoided the issue of the priority dispute with Quetelet, along with the contributions of others in France and England to what became the "moral statistics movement" in the middle and later part of the 19th century. Although his work was certainly honored during his lifetime, his modest personality and his position as a dedicated amateur without high academic or social connections left his accomplishments somewhat underappreciated outside of criminology.

## 3. MULTIVARIATE ANALYSES: DATA-CENTRIC DISPLAYS

As we have seen, Guerry's data consist of a large number of variables recorded for each of the départements of France in the 1820–1830s and therefore involve both multivariate and geographical aspects. In addition to historical interest, these data provide the opportunity to ask how modern methods of statistics, graphics, thematic cartography and geovisualization can shed further light on the questions he raised.

To put it another way, What could we do today, if Guerry arrived as a consulting client at our door and asked for assistance in understanding these data on moral statistics of France? This section and the following are merely meant to be suggestive of the kinds of exploratory, graphical methods that might be of use. To avoid going too far afield, I make little attempt at modeling these data here.

Along the way, I try to suggest some specific challenges and opportunities for further growth that these applications entail. More generally, these examples call for greater integration of multivariate statistical methods and data displays with spatially referenced analyses and displays. See Whitt, McMorris and Weaver (1997) and Whitt (2007) for some attempts to construct multivariate and spatial models addressing Guerry's data and questions.

The easiest approaches to the questions Guerry raised simply treat his data as a multivariate sample and apply standard analysis and visualization methods. In graphs we can represent geographic location by color coding or other visual attributes. This gives data-centric displays in which the multivariate data are shown directly and geographic relations appear indirectly. I do this in a spirit both of giving Guerry

However, the Franco-Prussion war began in July, 1870 and these papers may have been lost. Ironically, for the history of statistical graphics, Guerry's work was not the only casualty of this war: Charles Joseph Minard, whose works I have detailed elsewhere (Friendly (2002)), fled from Paris to Bordeaux in September 1870, "carrying only one light bag and some studies already begun" (Chevallier (1871), page 22, D. Finley, trans.); these also have been lost.

a helping hand and asking whether modern methods can shed any light on the questions that Guerry entertained. To provide others the opportunity to do the same and to issue this Guerry challenge publicly, the data from Guerry (1833) and other sources, and base maps of France in 1831 are provided at www.math.yorku.ca/SCS/Gallery/guerry/.

### 3.1 Bivariate Displays

Even simple scatterplots can be enhanced by showing more (or less) than just the data to aid interpretation or presentation of results. Figure 12 shows two versions of a plot of crimes against persons against literacy that Guerry might have found helpful in understanding and explaining the relation between them.

The left panel plots the data together with a 68% data (concentration) ellipse and a smoothed loess curve to focus attention on the overall relation between crime and literacy, and ask whether there is any indication that this is nonlinear. To highlight particular départements that might be unusual (as explained below), observations relatively far from the centroid are labeled. The right panel shows separate 68% data ellipses for each of the five regions of France together with ±1 standard error crosses for the centroid of each region.

The data ellipse (Monette (1990); Friendly (2007b)) of size $c$ is defined as the set of points $\boldsymbol{y}$ whose squared Mahalanobis distance $\leq c^2$,

$$D^2(\boldsymbol{y}) \equiv (\boldsymbol{y} - \bar{\boldsymbol{y}})^\top \boldsymbol{S}^{-1} (\boldsymbol{y} - \bar{\boldsymbol{y}}) \leq c^2,$$

where $\boldsymbol{S}$ is the sample variance–covariance matrix. When $\boldsymbol{y}$ is (at least approximately) bivariate normal, $D^2(\boldsymbol{y})$ has a large-sample $\chi^2_2$ distribution with 2 degrees of freedom, so taking $c^2 = \chi^2_2(0.68) = 2.28$ gives a "1 standard deviation bivariate ellipse," an analog of the standard univariate interval $\bar{y} \pm 1s$. The labeled points in Figure 12 are those for which $\Pr(\chi^2_2) < 0.10$. See Friendly (2006, 2007b) for details and extensions of these ideas.[11]

Thus, Guerry and his readers would have been able to see directly that overall there is no linear relation between crimes against persons and literacy, nor is there any hint of a nonlinear one. The départements that are labeled would have served to highlight the discussion along the lines that Guerry chose, for example, the Ariège is near the bottom on crime and also on literacy, while Indre stands about the same on literacy, but is near the top on personal crime.

The right panel, highlighting region differences, shows a number of things that Guerry did not observe (or comment on). For example, most of the variation between regions on these variables is due to the difference between the center and west vs. north and east on literacy, but these regional differences on crime are very small. The south of France stands out as being quite low on both of these variables, and, moreover, the within-region covariation is positive. Here, the Ariège does not stand out as being particularly unusual for the south of France.

As mentioned earlier, in his last work Guerry (1864) attempted to come to grips with aspects of the multivariate relations among moral variables. Were he around today, he would surely find scatterplot matrices of interest, so we have prepared one for his inspection. Figure 13 shows a number of interesting relations: There is in fact a moderately positive relationship between personal crime and property crime for France as a whole, apparently contradicting his claim that they are negatively related. There are also negative, and possibly nonlinear, relations between literacy ($x$) and property crime, suicides, and children born out of wedlock, and so forth. The negative relation between suicide and personal crime that Guerry claimed is there, but rather weak; he might also be surprised by the positive relation between suicide and property crime.

However, Figure 13, like any other graph, shows some features of the data and hides others. In particular, as we saw in Figure 12, there are large regional differences in some of these measures which help to explain some apparent findings for the total sample.

### 3.2 Reduced-Rank Displays

Visual summaries, such as the data ellipse and loess smooths, used within a scatterplot as in Figure 13, may show quite effectively the relations among a reasonably large number of variables. Yet there are even better methods for display of complex, high-dimensional data.

[11] In particular, one should note that these normal-theory (first and second moments) summaries can be distorted by multivariate outliers, particularly in smaller samples. In principle, such effects can be countered by using robust covariance estimates such as multivariate trimming (Gnanadesikan and Kettenring (1972)) or the high-breakdown bound minimum volume ellipsoid (MVE) and minimum covariance determinant (MCD) methods developed by Rousseeuw and others (Rousseeuw and Leroy (1987); Rousseeuw and Van Driessen (1999)). Often, these robust methods supply weights that can be used to "robustify" other multivariate methods and visualization techniques, but this integration in applied software is still quite spotty.

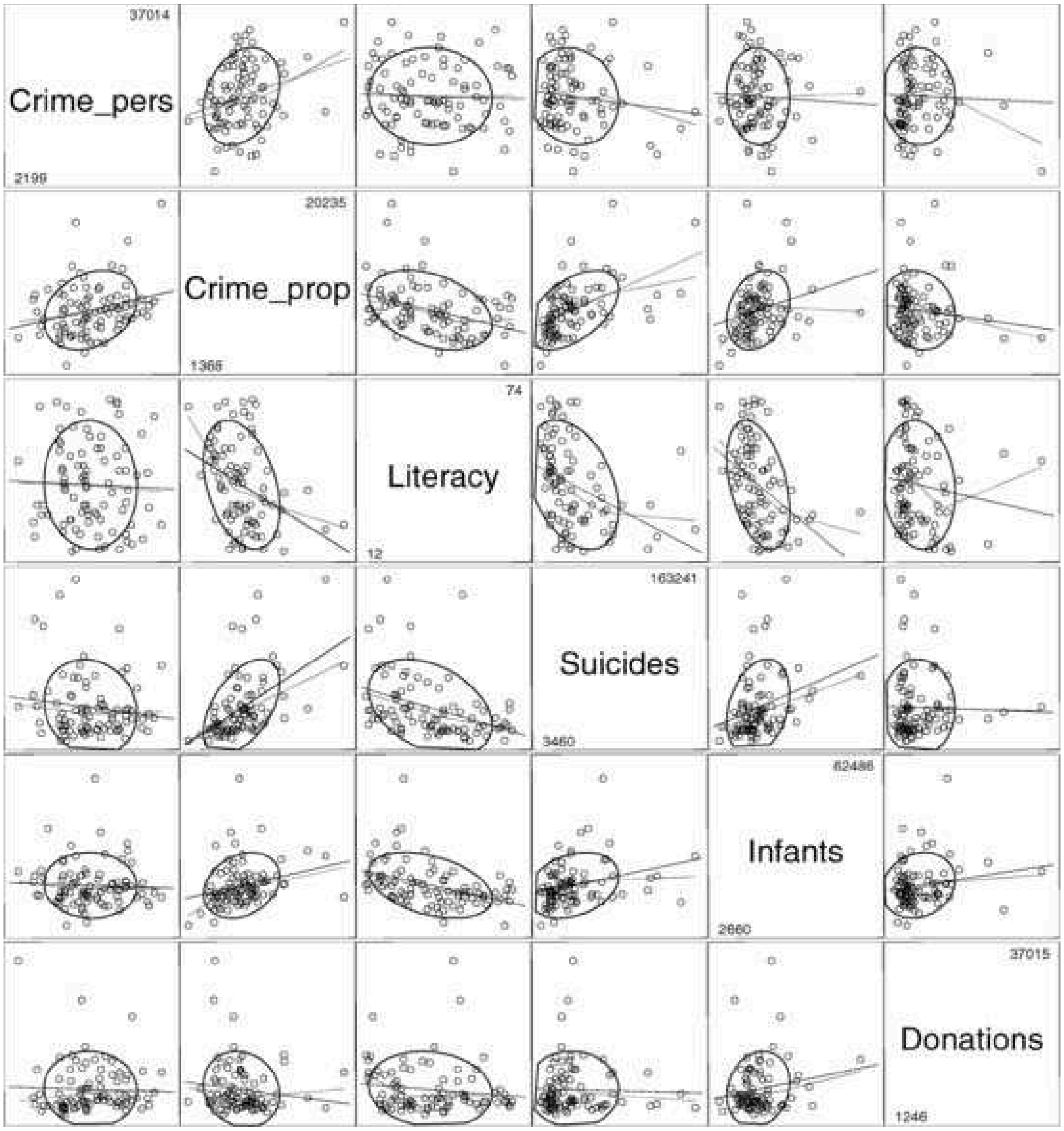

FIG. 13. *Enhanced scatterplot matrix for crimes against persons, crimes against property, literacy, suicides, children born out of wedlock (infants naturelles) and donations to the poor. Each panel plots the row variable on the vertical against the column variable on the horizontal.*

3.2.1 *Biplots.* Among these, the biplot (Gabriel (1971), 1981) must rank among the most generally useful. Biplots can be regarded as the multivariate analog of scatterplots (Gower and Hand (1996)), obtained by projecting a multivariate sample into a low-dimensional space (typically of 2 or 3 dimensions) accounting for the greatest variance in the data. The (symmetric) scaling of the biplot used here is equivalent to a plot of principal component scores for the observations (shown as points), together with principal component coefficients for the variables (shown as vectors) in the same 2D (or 3D) space. When there are classification variables dividing the observations into groups, we may also overlay data ellipses for the scores to provide a low-dimensional visual summary of differences among groups in means and covariance matrices.

For example, Figure 14 shows a 2D biplot of Guerry's six quantitative variables. In this plot, (a) the variable vectors have their origin at the mean on each variable and point in the direction of positive deviations from the mean on each variable (more is better); (b) the angles between variable vectors portray the correlations between them, in the sense that the cosine of the angle between any two variable vectors approximates the correlation between those variables (in the reduced space); (c) the orthogonal projections of the observation points on the variable vectors show approximately the value of each observation on each variable. (These interpretations assume that the axes are equated, so that a unit data length has the same physical length on both axes in the plot.)

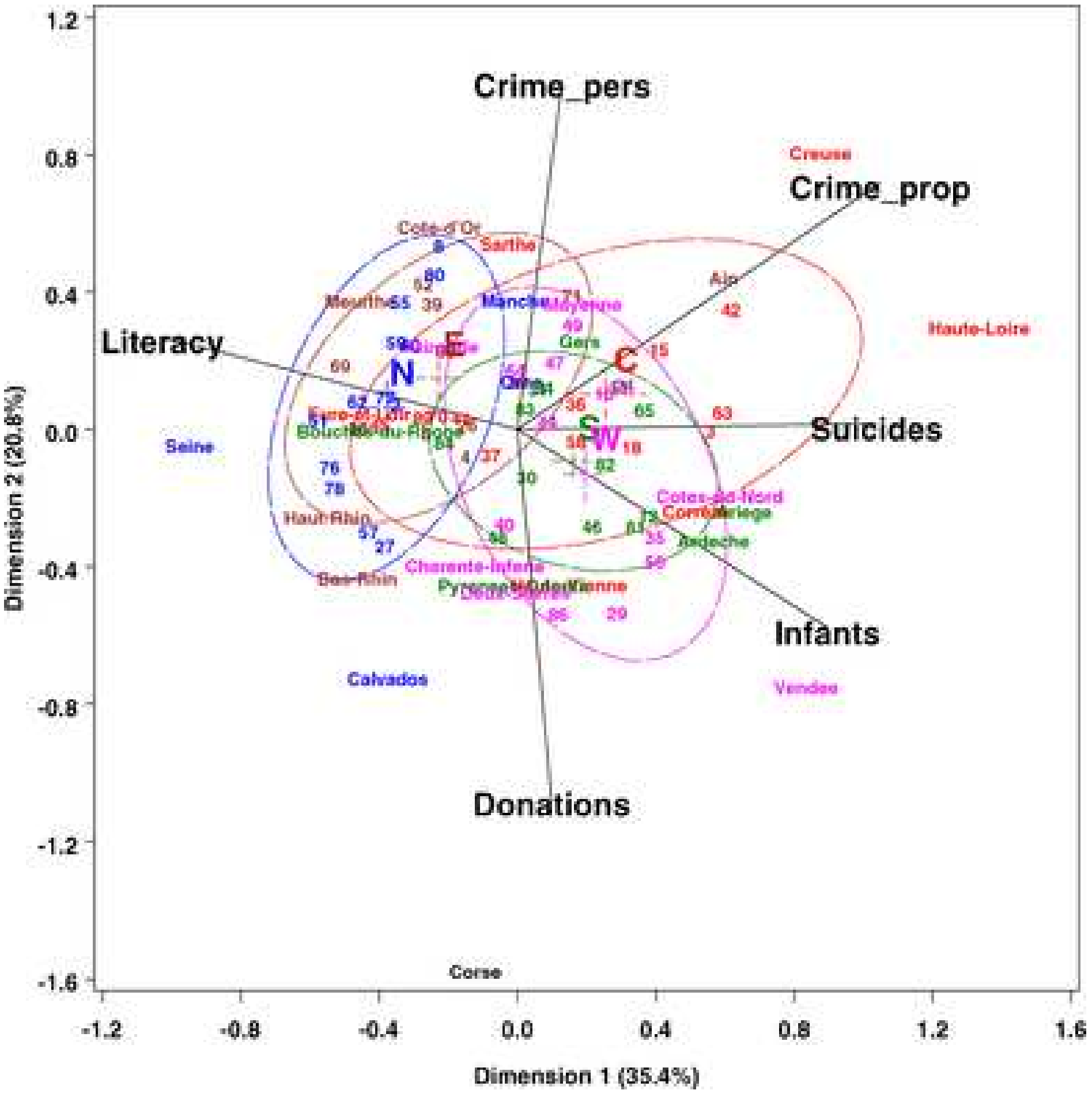

FIG. 14. *Biplot of Guerry's six quantitative moral variables, with 68% data ellipses for regions, dimensions 1 and 2. To avoid overplotting, départements within their ellipse are labeled by number rather than name.*

The two dimensions account for 35.4% and 20.8%, respectively, of the total variance of all measures. The total, 56.2%, is relatively low, reflecting the generally small correlations among these moral variables. Nevertheless, a number of interesting features may be read from this plot. The first dimension, aligned mainly with literacy on one side and suicides on the other is largely the distinction between France obscure and France éclairée, with the North and East higher on literacy than the other regions. Seine (consisting essentially of Paris) stands out as particularly high on literacy, while Creuse, Haute-Loire and Vendée are particularly high on property crime, suicides and *infants naturelles*.

The second dimension might be described as "benevolence," contrasting personal crime against donations to the poor, with Corsica standing out as the most extreme on both. Guerry does not suggest any comfort from the knowledge that getting robbed in Corsica might be offset by gifts to charitable establishments, but he does note from his map that most of the donations to the poor are found to the southeast of a line from Côte d'Or to Ariège and that "if Corsica is excluded, one encounters the greatest contributions to the poor in those départements where the Catholic clergy is most widespread and where crimes against persons are at the same time most common" Guerry (1833, page 114, WR trans.).

3.2.2 *Canonical discriminant plots.* If instead of accounting for variation among all the départements, we ask what low-rank view shows the greatest differences among regions of France, we are led to canonical discriminant (CD) plots ((Friendly, 1991, Section 9.5.2)). This method plots scores for the départements on linear combinations of the variables which maximally discriminate among the group mean vectors, in the sense of giving the largest possible univariate $F$ statistic. It is equivalent to a canonical correlation analysis between the set of response (moral) variables and a set of dummy variables representing group membership (region).

The CD plot for Guerry's data is shown in Figure 15. Similar to the biplot, we have added variable vectors whose angles with the axes indicate the correlation of each variable with the canonical dimensions to aid interpretation. The length of each variable vector is then proportional to its contribution to discriminating among the means for the regions of France. Because the canonical variates are uncorrelated, confidence ellipses for the means of each Region plot as circles (Seber (1984)).

Here we can account for over 90% of the differences in the means of regions in two dimensions. The first dimension is most highly correlated with literacy and suicide, but the latter distinguishes most among the regions. The configuration of the regions largely reflects France obscure vs. France éclairée. Along the second dimension, the means for region are correlated so that donations to the poor and personal crime correlate positively, and the south of France differs most from the west.

3.2.3 *Graphical challenge: labeled scatterplots.* In a number of historical studies and reviews (Friendly (2005); Friendly and Denis (2005); Friendly (2007a)) I have been struck by what may be learned in the process of working with old data in a modern context, a topic I call *statistical historiography.* In this connection it is useful to comment on one specific challenge faced here in attempting to create the plots shown so far in this section with modern software.

Although Guerry was, throughout his career, fundamentally concerned with *relations* among moral variables, he never thought to draw a scatterplot, because this graphical form as such was not invented until 1833 (Herschel (1833)) and did not enter common use until after he had died (Friendly and Denis (2005)). But he always wanted to interpret findings in relation to the physical and social geography of France. The identity of départements in the map of France was at least implicitly known to Guerry's readers; in any case, he provided a legend. In a scatterplot, he would have wanted to identify the points and would have carefully positioned the names by hand to maximize legibility.

In Figure 12, I chose to label only those points outside the 90% data ellipse to highlight those départements that were atypical in the relation between property crime and literacy. In a static graph, this must be done by programming; interactive graphic software allows identification by mouse clicks. Perhaps you did not notice what was missing, but Guerry certainly would have.

The initial version of the biplot (Figure 14) labeled all points with the département name, but many of these were obscured due to overplotting. In the version shown here, the labels were thinned by replacing names by département numbers within the data ellipse for each region; this is better visually, but requires lookup (see the Appendix).

The CD plot in Figure 15 was a greater challenge, because I felt it was important to identify all the départements directly. In the end, I carefully adjusted graphic parameters, but then hand-edited the PostScript figure to reduce occlusion.

There have been a variety of proposals for more automatic ways to label a maximum number of points in a plane (scatterplot or map) with minimum overlap, in the psychometric and statistical graphic literature (e.g., Kuhfeld (1991), 1994; Noma (1987)), in cartography (e.g., Hirsch (1982)), computer science and graphic information systems (e.g., van Kreveld, Strijk and Wolff (1999)). An extensive map-labeling bibliography (Wolff and Strijk (1996)) now lists over 300 references on this topic. As far as I know, none of these is implemented in commonly available statistical software, except for Kuhfeld (1994) in SAS and the R function `thigmaphobe.labels` in the plotrix package (Lemon et al. (2007)).

## 3.3 HE plots for Multivariate Linear Models

I close this section with one more display that relates to Guerry's aspirations in 1864 to address the multivariate relations of crime with other moral and population characteristics. Figure 16 is based on a scatterplot of crimes against persons and property corresponding to the panel in row 1, column 2 of Figure 13, with points outside a 95% data ellipse identified by département name.

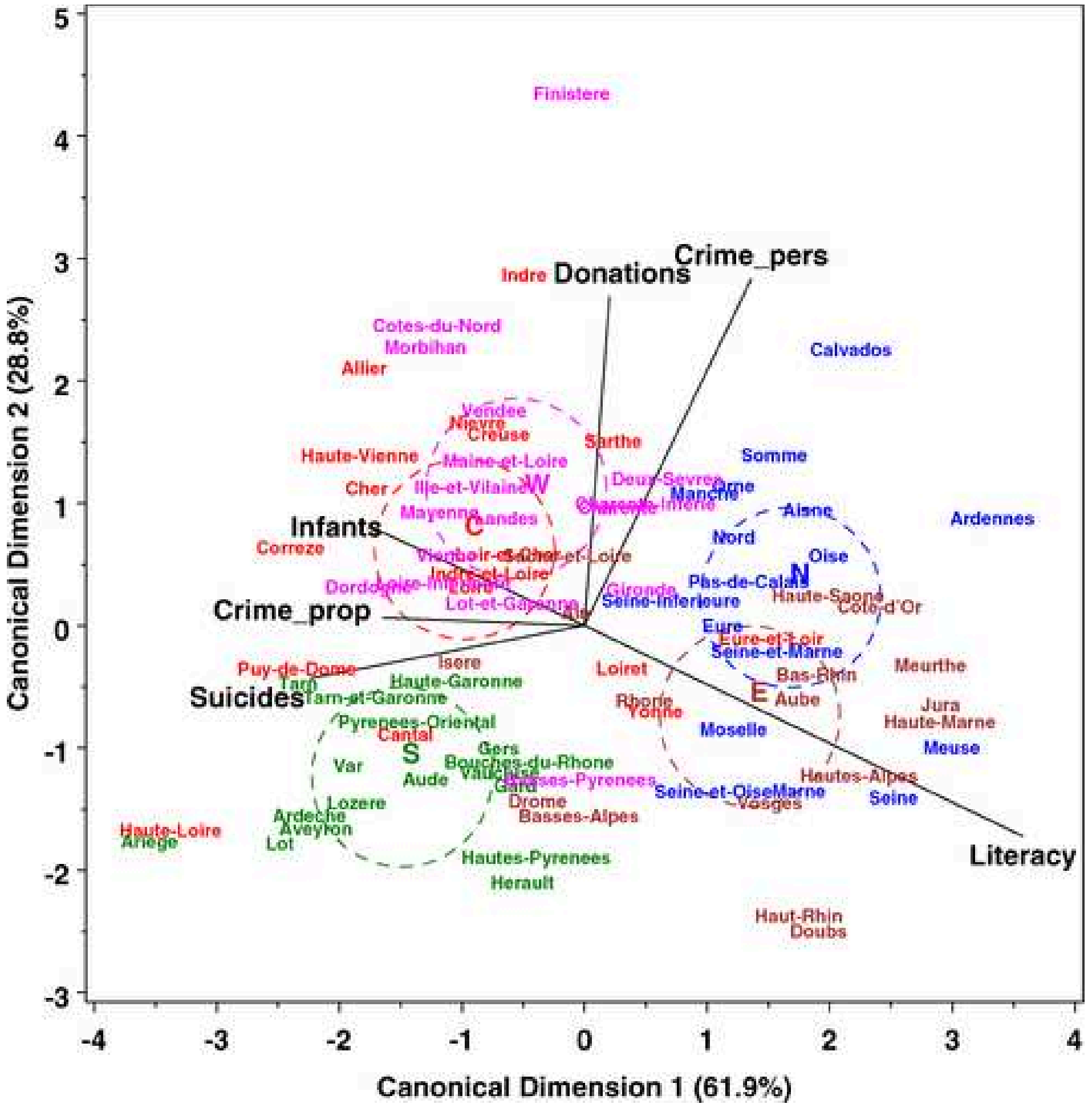

FIG. 15. *Canonical discriminant plot for Guerry's six quantitative variables. Circles show 99% confidence regions for the region means; variable vectors indicate the correlation of each variable with the canonical dimensions.*

This is overlaid with an hypothesis–error (HE) plot (Friendly (2006), 2007b; Fox et al. (2007)) that provides a compact visual summary of the hypothesis (H) and error (E) covariation in the multivariate linear model, $\boldsymbol{Y} = \boldsymbol{XB} + \boldsymbol{E}$. Here we are fitting population per crime against property and persons to the region factor and the quantitative effects of suicides, literacy, donations, infants and wealth.[12] In R/S-Plus notation, this is

```
guerry.mod
  <- lm( cbind(Crime_prop,
               Crime_pers)
         ~ Region + Suicides
         + Literacy + Donations
         + Infants + Wealth)
```

This model yields an $R^2$ of 0.43 for property crime and 0.36 for personal crime, and gives the multivariate analysis of variance (MANOVA) test statistics below. The HE plot provides a direct visualization of the size and nature of these effects.

In Figure 16, the error ellipse is the 68% data ellipse of the bivariate residuals ($\boldsymbol{E}$) divided by the error degrees of freedom $(n - p)$ and centered at

[12] I am using this simply as an example here, rather than a full-blown attempt to model crime in relation to all of Guerry's moral variables. In particular, it can be argued that crimes should be expressed as the reciprocals of Guerry's measures, giving rates per unit population as is commonly done today. Whitt, McMorris and Weaver (1997) provided analyses of a series of univariate linear models for most of Guerry's moral variables and discussed these in relation to social theory and Guerry's findings. As in the present example, these use ordinary least squares methods which ignore the spatial autocorrelation (e.g., Anselin and Bera (1998)) of residuals surely present in such geographical data. A recent paper by Whitt (2007) applied a variety of spatial regression models and visualizations to examine the relation between crimes of violence and suicides. Yet another challenge is the extension of these methods to multivariate linear models.

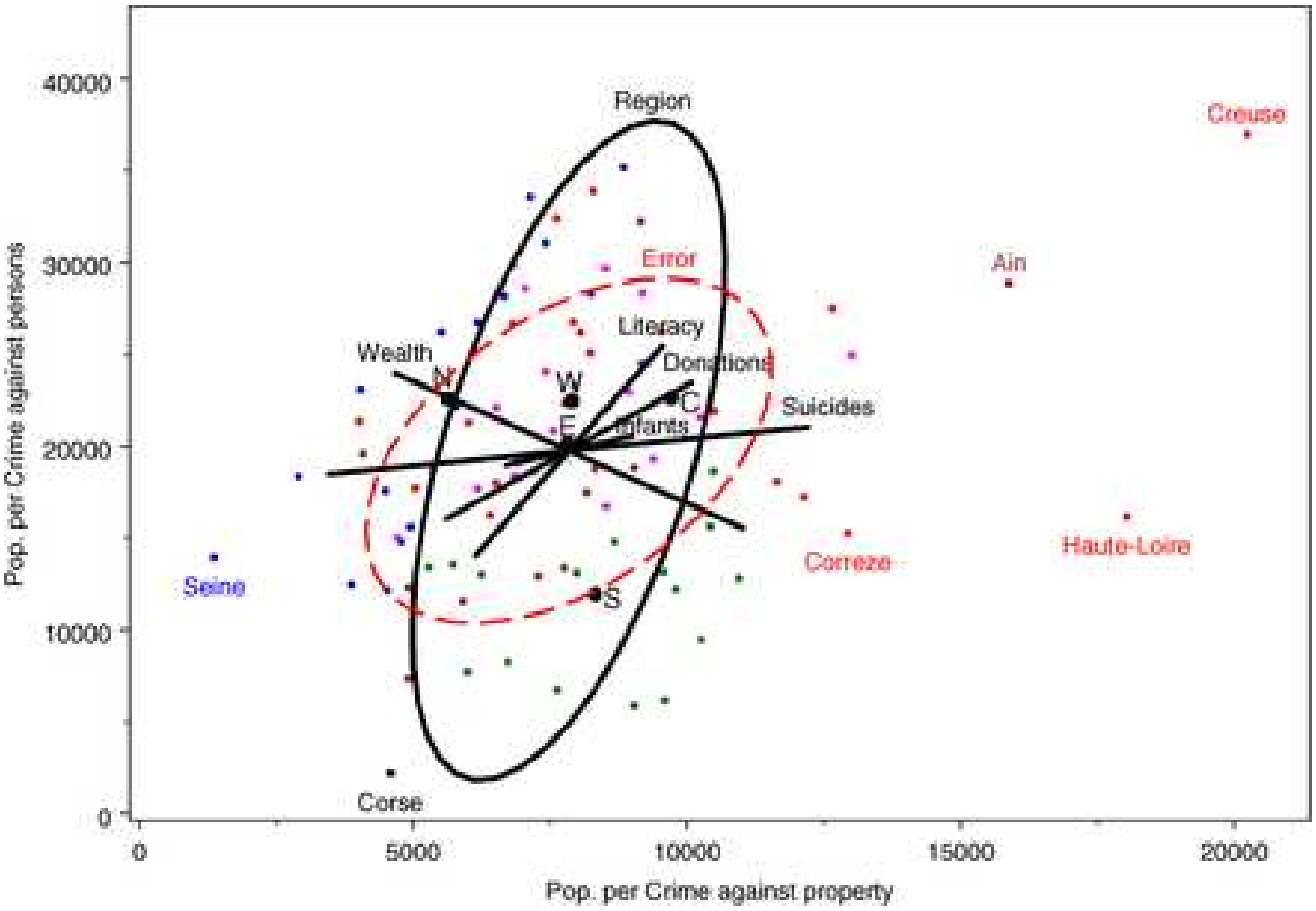

FIG. 16. *HE plot for the multivariate linear model fitting rates of crime to region and the covariates suicides, literacy, donations, infants and wealth. The black, solid ellipses (and degenerate lines) show the (co)variation of the hypothesized predictor effects relative to the red, dashed error (E) ellipse. They have the property that they protrude beyond the E ellipse* iff *the hypothesis can be rejected by the Roy maximum root test.*

```
> Manova(guerry.mod, test="Roy")

Type II MANOVA Tests: Roy test statistic
          Df test   stat   approx F  num Df  den Df     Pr(>F)
Region       5     0.6859    10.2878      5      75   1.554e-07 ***
Suicides     1     0.1437     5.3170      2      74   0.006957 **
Literacy     1     0.0361     1.3354      2      74   0.269328
Donations    1     0.0336     1.2444      2      74   0.294059
Infants      1     0.0091     0.3385      2      74   0.713923
Wealth       1     0.1479     5.4719      2      74   0.006077 **
---
Signif. codes:  0 '***'  0.001 '**' 0.01 '*' 0.05 '.' 0.1 ' ' 1
```

the grand means, to put it on the same scale as the data. The E ellipse thus represents the partial covariation between the crime rates, controlling for the model effects. The orientation of the E ellipse indicates that crimes against persons and property are still positively related after adjusting for those effects. Its shadows on the axes are proportional to the residual standard errors on the data scale.

Each predictor effect is shown as the bivariate H ellipse for the Type II sum of squares and cross-products matrix ($\mathbf{SSP}_H$) used in the corresponding multivariate test. For a 1 df hypothesis, such as the five quantitative regressors, the $\mathbf{SSP}_H$ is of rank 1 and the H "ellipse" collapses to a line. The H ellipses have all been scaled to protrude outside the E ellipse *iff* the corresponding Roy maximum root test is significant at a conventional $\alpha = 0.05$ level. [This scaling is produced by dividing $\mathbf{SSP}_H$ by $\lambda_\alpha(n-p)$, where $\lambda_\alpha$ is the critical value of Roy's statistic for a test at level $\alpha$.] Thus, the directions in which the hypothesis ellipse exceed the error ellipse are informative about how the responses depart significantly from $H_0$.

Quite a lot can be read from this plot. It is far less detailed (and hopefully more comprehensible) than Guerry's (1864) *General Causes of Crimes* (Figure 11), but attempts to address similar issues of multivariate relations. Region differences are particularly large in personal crime, but not so in property crime (controlling for the regressors). Suicide and wealth (a ranked index based on taxes on personal and movable property per inhabitant) are strongly related to crimes against property, but not to crimes against persons. In this model literacy, donations to the poor and infants are not individually significant predictors of crime, although their predicted effects are positively related to each other. The variation in the region means on the crime variables may be read directly, as noted earlier, and the six atypical départements identified by point labels can also be understood in relation to the plots we have seen above.

## 4. MULTIVARIATE MAPPING: MAP-CENTRIC DISPLAYS

Compared with the data-centric displays just reviewed, the cartographic display of multiple phenomena simultaneously, called *multivariate mapping*, has the advantage of preserving geospatial context, but creates its own difficulties and challenges. These include: (a) how to encode multiple variables in a geographic unit for different purposes (readability, exploring relations among variates, detecting unusual patterns); (b) how to relate maps to model-based summaries; (c) how to show indicators of variability, uncertainty or data quality. I illustrate just a few methods below,[13] simply to illustrate some current techniques, discuss their limitations and encourage others to do better.

### 4.1 Star Maps

Multiple variables can be displayed together on a single map in a variety of ways (e.g., Slocum et al. (2005)) but most of these fall into two main categories: either the separate variables are assigned to different visual attributes of a glyph or point symbol [e.g., angle and length of rays, height and width of rectangles, facial features in faces symbols (Chernoff (1973)) and so forth], or they are combined into more integral forms, such as when two variables are represented by shades of two complementary colors overlayed (e.g., by additive or subtractive mixing) to determine the shading color of each region (Trumbo (1981)). Among the former class, ray glyphs or star symbols (radial lines positioned around a circle, each of length proportional to a given variable) are widely used and can be used for any number of variables.

Figure 17 shows two versions of a star map for Guerry's main variables designed for different purposes. In graphics it is always true that details matter, but this is particularly true for displays like this, where there are many choices, not all of them obvious. The order of the variables around the circle greatly affects the kinds of shapes that appear. Here, we want to use the information about correlations among variables to simplify the shapes, so the variables were ordered according to their angular positions in the biplot (Figure 14), following the principle of correlation effect ordering (Friendly and Kwan (2003)). Second, scaling and orientation of the variables matter in determining the relative size and shape of the stars. Here, Guerry's variables, scaled so that larger numbers reflect better outcomes, were first converted to ranks and the ranks were assigned to the lengths of the rays in such a way that bigger also corresponds to a better or more moral result on each variable. Thus, départements with better outcomes on all measures are shown as larger stars. If one wanted to focus on the départements with worse outcomes, encoding ray length by reverse rank would be more appropriate.

Finally, the star glyphs may be colored to reflect some other aspect or extended in other ways. In Figure 17 we focus on the comparisons across regions of France. The figure at the right shows a schematic summary of the distributions within each region using overlaid star glyphs to show the median, lower quartile (white) and upper quartile (grey), a kind of simple multivariate boxplot glyph (showing only the middle 50% of each distribution). Alternatively, if we want to capture statistical characteristics of the profile for each département, we might use color to encode the mean moral rank (reflected by size of each glyph) or the standard deviation of the ranks (reflected by eccentricity of the shape), as is done in Figure 18.

In such figures, it will be seen that it is the configural properties of size and shape that attract the eye; individual variables can be read, with a bit of effort and aid of the variable key. Figure 17(b) provides a relatively compact summary of the differences among regions on all variables simultaneously. In the plots showing individual départements, Corsica stands out because it is very bad on crime, but relatively high on other variables; in the north, regions around Paris [Seine (75), Yvelines (78) and Marne (51)] are all small and similar in shape, reflecting high literacy, but relatively low on all other variables; in the east, most départements are relatively large, with the exception of Drôme (69); in the center, most départements are relatively good on all variables except literacy; in the south, most départements are relatively poor on a number of variables, particularly donations to the poor, and Bouches-du-Rhone (13), whose main city is Marseilles, and Lozère (48) stand out as particularly low. These displays are necessarily quite complex, but they package a lot of information. It would be of interest to know what Guerry would make of them.

[13] The département and region boundaries for the base map used in these displays were created starting with a modern map of France, with départements merged or deleted to adjust it to that in 1830. The map files in various formats are available on the companion web site for this article. Département names, numbers and regions are listed in the Appendix.

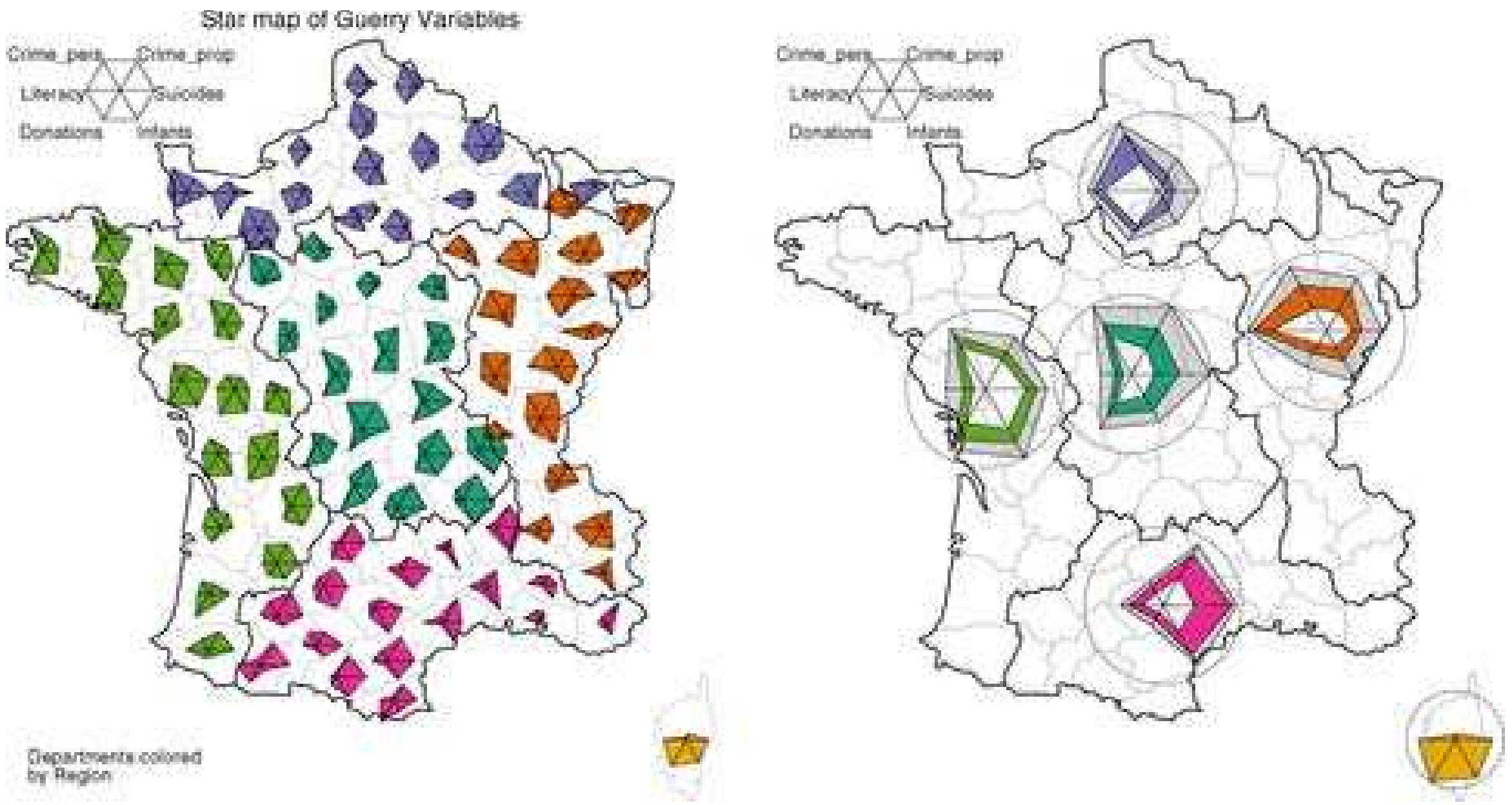

Fig. 17. *Star maps of Guerry's data, using rays proportional to the rank of each variable (longer = better). Variables have been ordered according to their angular positions in the biplot (Figure 14). Left: Glyphs for individual départements. Right: Multivariate boxplot glyphs for the medians and quartiles across départements in each region of France.*

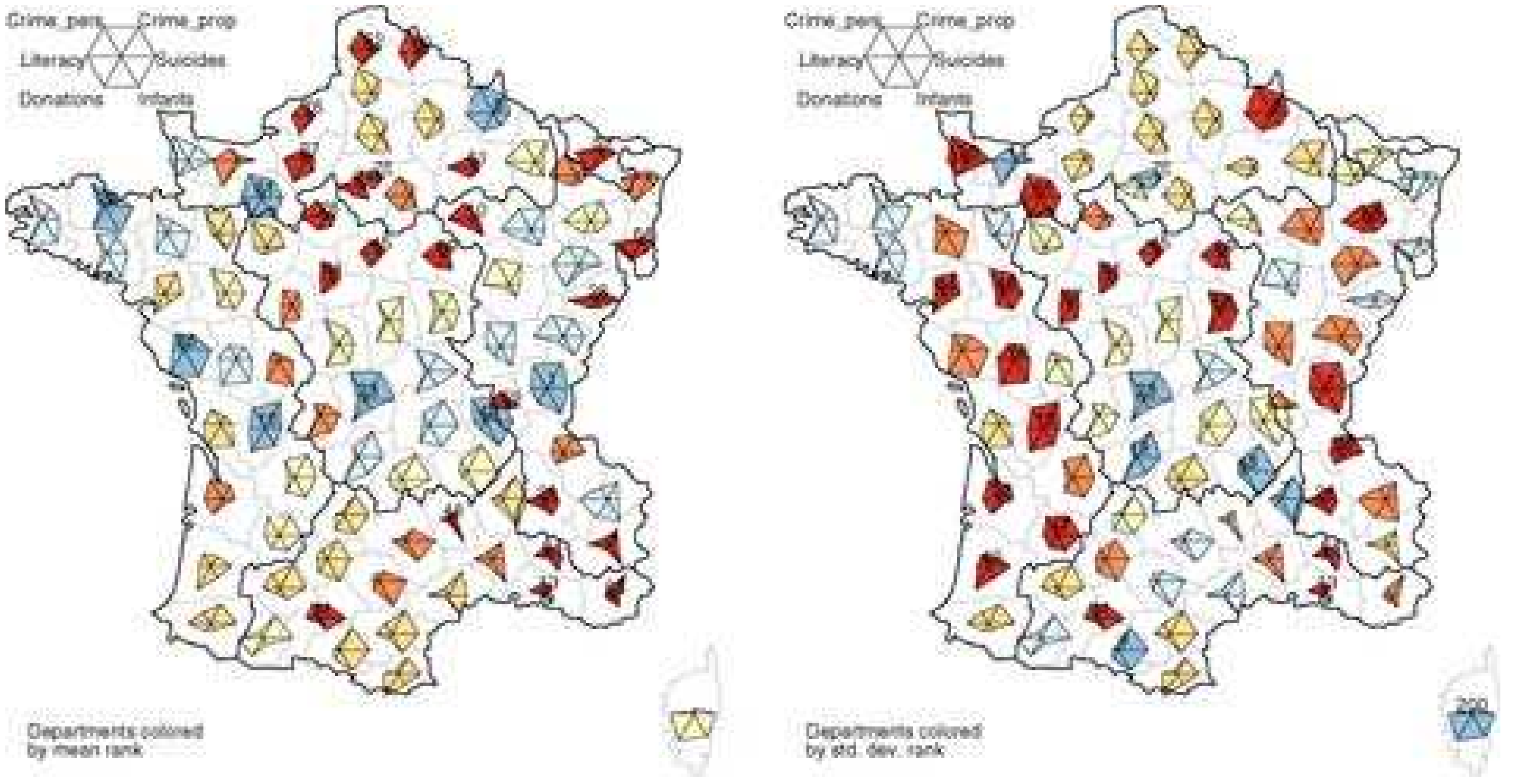

Fig. 18. *Using color for other encodings. Left: Mean rank; right: standard deviation of the ranks. Départements that are unusually high or low on each measure are identified by département number.*

### 4.2 Red–Green–Blue Blended Color Maps

A variety of schemes can be used to combine two or more variables in bi-, tri- or multivariate choropleth maps in more integral ways. When color is used, it is most convenient, both conceptually and computationally, to use various methods to blend or interpolate composite colors in red–green–blue (RGB) color space, even though it is well known that RGB colors are neither perceptually uniform (e.g., in brightness) nor perceptually linear. For the present purposes, the simplicity of RGB blending is sufficient to convey the main ideas. Thus, for three variables, $x_1, x_2, x_3$, we use the color mapping function, $\mathcal{C}(x_1, x_2, x_3) \mapsto \text{rgb}(x_i, x_j, x_k)$ for $i, j, k$ some permutation of 1, 2, 3. See Ihaka (2003) for a discussion of alternative color spaces from a statistical perspective.

Figure 19 shows a trivariate RGB map of crimes against persons, crimes against property and literacy, using a linear scale to map each variable into 0–100% of the color indicated in the legend.[14] The color combinations of the départements form a rather wide range of the spectrum and are relatively easy to interpret, remembering that more is better for each variable. Thus, départements where literacy is high are shaded in blues and purples, and more blue to the extent that crime rates are low; the north of France is primarily in this category. Those that have high values for both population per crime variables (low crime rate) but are low in literacy are colored yellow, such as Creuse (23) and Ain (1). Corsica, with relatively low values for population per crimes (high crime rates), but moderately high literacy is shaded bluish.

[14]It is difficult to indicate all possible three-variable color blends compactly. The actual colors plotted correspond to color blends calculated continuously by linear interpolation. The legend in Figure 19 uses trilinear coordinates to show the colors corresponding to *relative* amounts of red, green and blue; thus vectors of (0.25, 0.25, 0) and (0.75, 0.75, 0) both map to yellow in the legend, but appear differently in the plot.

This scheme for color-blended choropleth maps may be extended to more than three variables by use of rank-reduction techniques. One simple idea is to use principal components or factor analysis to obtain scores for each department on three components or factors, to which color blending is applied. Table 1 shows the result of a principal component analysis followed by varimax rotation for Guerry's six moral variables, with tentative labels for three factors.[15]

Applying these weights to the variables gives scores for each départment on the three uncorrelated dimensions, $F_1, F_2, F_3$. Thus, $F_1$ will be large when the *rates* of property crime, illegitimate births and suicides are low; from Figure 13 it may be seen that these are all inversely related to literacy. $F_2$ relates positively mainly to donations to the poor, and $F_3$ is an index of crime, most heavily weighted on personal crime.

The color mapping function $\mathcal{C}(F_1, F_2, F_3) \mapsto \text{rgb}(F_1, F_2, F_3)$ then produces Figure 20. The north–south distinction is again evident, with shades of blue in most of the north going toward reds and purples to the south. A number of départements stand out in relation to their neighbors. These are all départements that appear as outliers for their regions in the biplot (Figure 14): Calvados (14), Corsica (200), Creuse (23), Haute-Loire (43) and Vendée (85).

### 4.3 Conditioned Choropleth Maps

In spatial data analysis, interest is often focused on one or two main geographic indicators, but it is desired to see how this changes or varies spatially when other, background variables are controlled or accounted for. This is certainly true for Guerry's data, where the main focus is on rates of crime and one would like to control for potential predictors such as literacy or economic conditions.

[15]In an unrotated principal components solution, the weights for variables are identical to those shown in the biplot (Figure 14).

TABLE 1
*Rotated component loadings for three-factor RGB blending; component weights less than 0.30 are not shown*

| Variable | Factor 1 Civil society | Factor 2 Moral values | Factor 3 Crime |
|---|---|---|---|
| Pop. per crime against persons | | | 0.97 |
| Pop. per crime against property | 0.75 | | 0.39 |
| Percent read & write | −0.72 | | |
| Pop. per illegitimate birth | 0.62 | 0.42 | |
| Donations to the poor | | 0.89 | |
| Pop. per suicide | 0.80. | | |

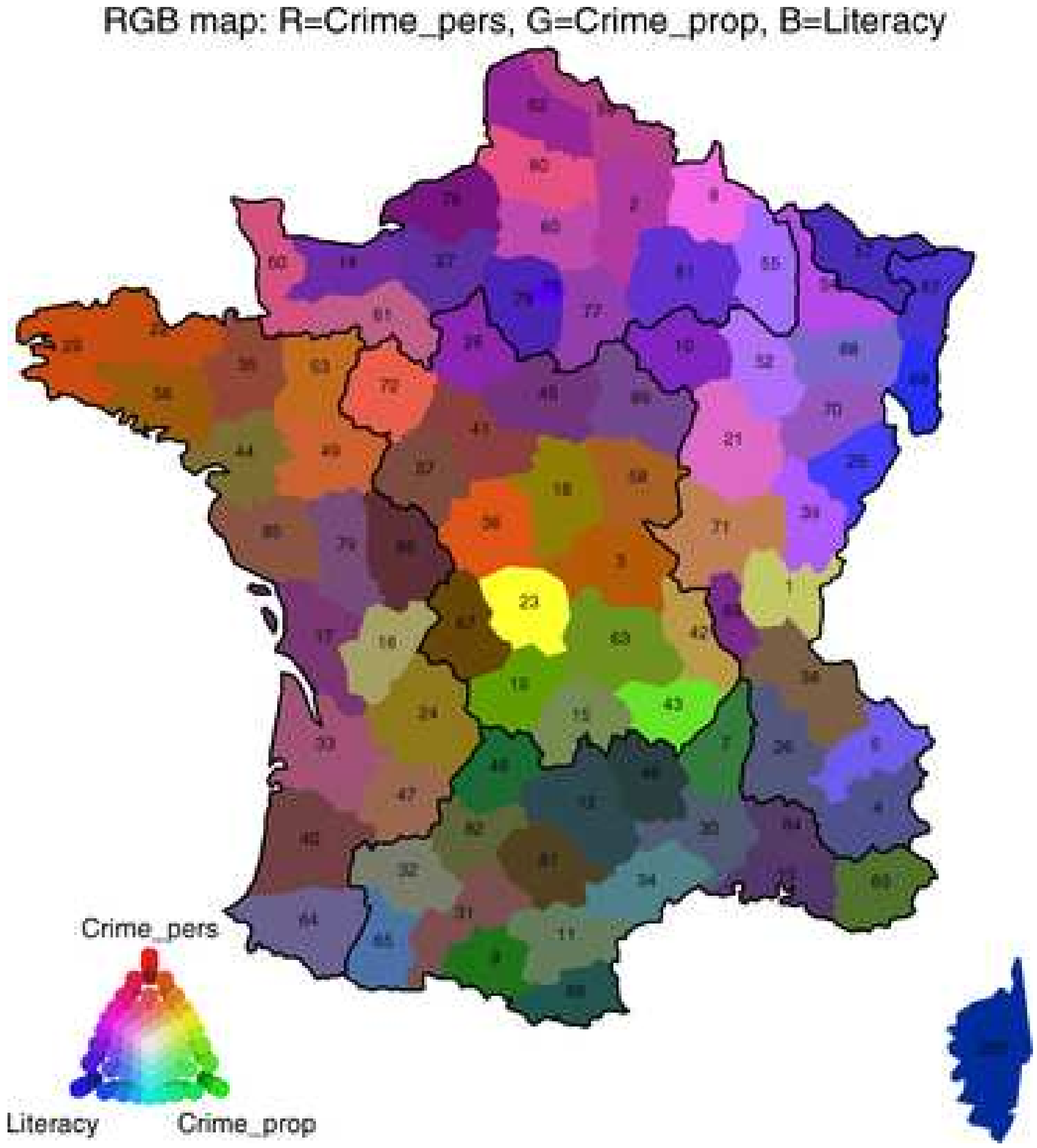

FIG. 19. *RGB map of crimes against persons, crimes against property and literacy. The legend shows the mapping of colors to the variables and the combinations that result from RGB blending in trilinear coordinates (indicating the* relative *amounts of R, G and B).*

One conceptually simple approach is to fit a model using the background variables as predictors and then display the residuals on a map. Then, however, the background variables are removed from the map display as well as from the response; we see only the deviation of the variable of interest from its expectation under a particular model. An attractive alternative is to use the predictors as conditioning variables, splitting the geographic units into subsets and showing the resulting set of maps in a coherent multipanel display. This method, called a *conditioned choropleth map* (Carr, White and MacEachren (2005)), thus provides a map-based analog of coplots or trellis displays (Cleveland, Grosse and Shu (1992); Cleveland (1993)) that have proved useful for similar graphic analysis of nonspatial data.

In a basic conditioned choropleth map (CC map), two potential predictor ("given") variables, $x$ and $y$, can be used to allocate the geographic regions (départements, here) into cells in a two-way table of size $n_x \times n_y$ using either nonoverlapping or overlapping ranges (called *shingles*). The CC map is then an $n_x \times n_y$ array of choropleth maps of the response variable for just those regions in each cell; for context, other regions are also shown, but with a background color. An interactive implementation of CC maps by Dan Carr and Yuguang Zhang (www.galaxy.gmu.edu/~dcarr/ccmaps) provides dy-namic selection of the conditioning intervals and the color coding of the response levels via sliders.

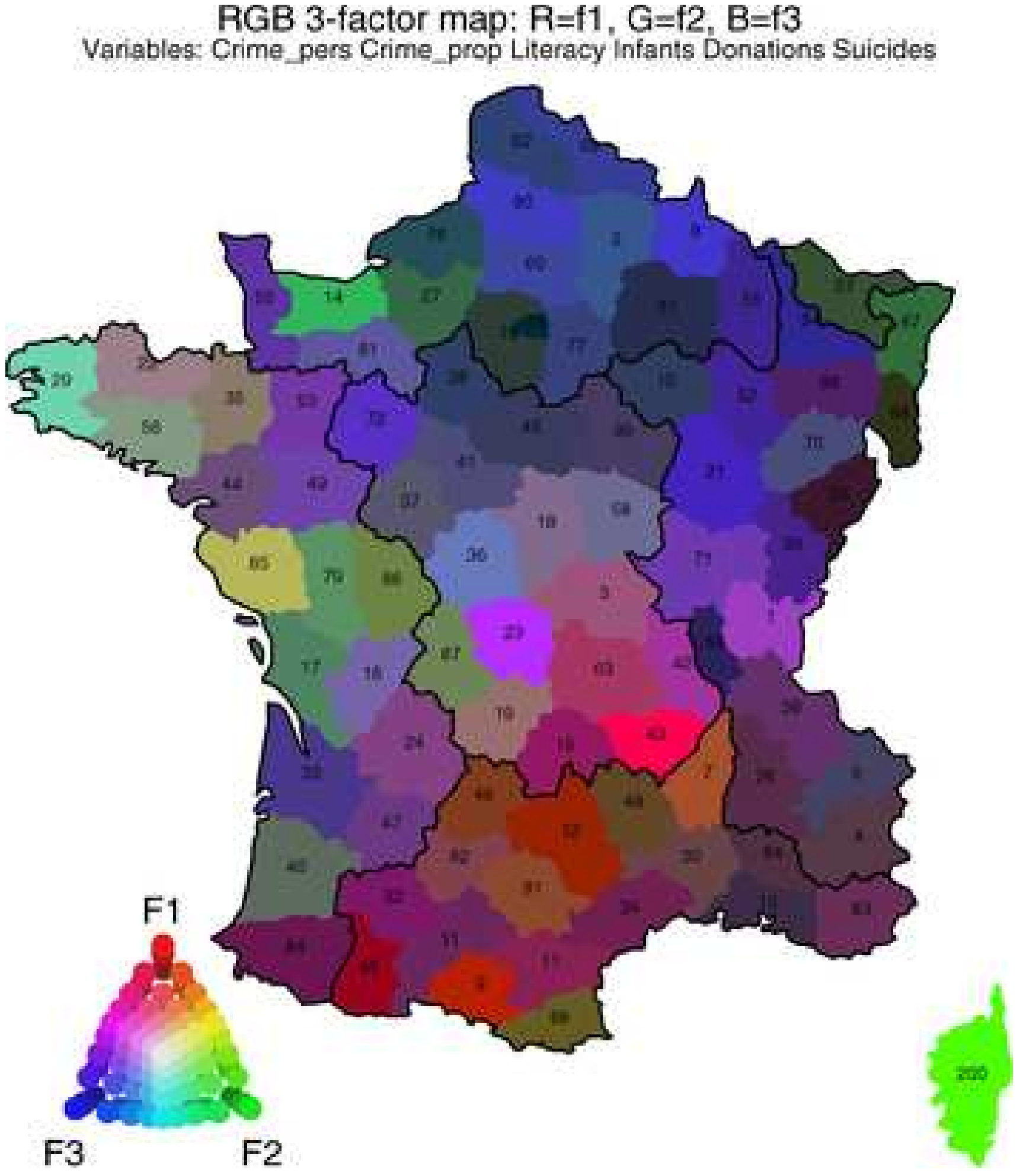

Fig. 20. *Reduced-rank RGB map of Guerry's main moral variables.*

Figure 21 shows an example of a static display of property crime, conditioned on literacy and economic wealth (measured as the rank order of taxes on personal property per inhabitant, where 1 is the maximum, 86 is the minimum). Both conditioning variables were divided into two shingles, allowing 10% overlap for each. Crime rate, expressed as population per personal crime, is shown by a bipolar (diverging) color scheme based on cutting the complete distribution of personal crime at percentiles from 20 to 80 in steps of 10 and using shades of red or blue to denote, respectively, higher or lower levels of crime.[16] For each panel, the marginal yellow bars show the values of literacy and wealth used in that conditional map, with lengths proportional to the percent of départements falling into each interval. Annotations in the upper left corner of the panel show the median crime rate and number of départements represented in that panel.

The interpretation of Figure 21 is as follows. The upper right panel shows the départements where literacy is high and wealth is high (small ranks); except for a few, these are mainly in the north of France. But, as Guerry concluded, these are, paradoxically, largely the départements in which there are the greatest numbers of crimes, as can be seen by the number shaded red. Conversely, the lower left panel shows the départements where literacy is low and wealth is low, which are all found in France ob-scure. Here, rate of property crime is relatively low, particularly in the center of France (shaded blue).

[16] Although crime is a unipolar variable, it seems more useful to use a diverging bipolar color scale here to focus attention on regions that are higher or lower than average. Use of red for "high risk" vs. blue for "low risk" is conventional in areas where CC maps are used.

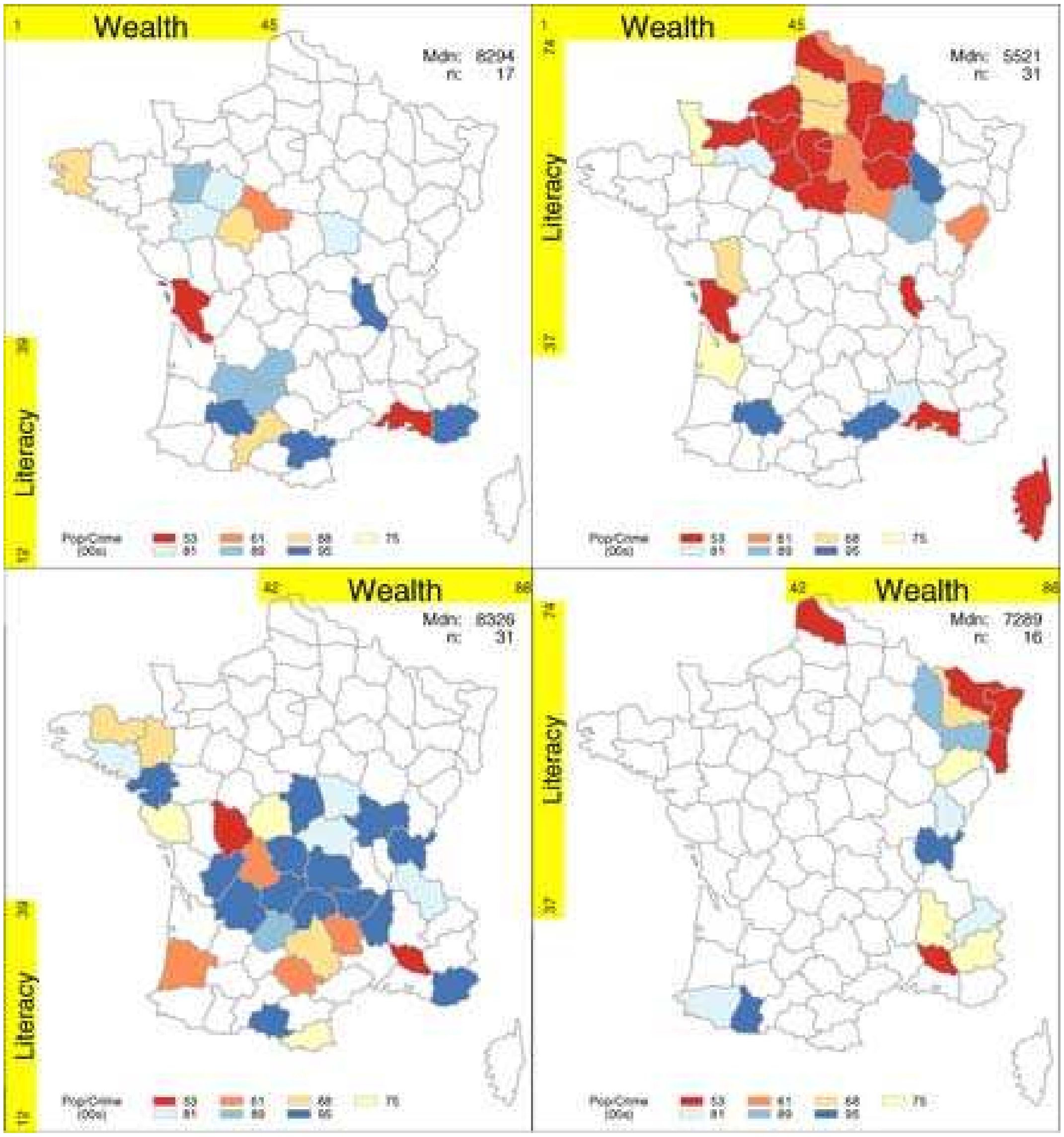

FIG. 21. *Conditioned choropleth map for population per crime against property, stratified by literacy (%) and wealth (rank, 1 = best). For each quadrant, only those départements in the indicated ranges of literacy and wealth are shaded. Shading levels correspond to percentiles of crime, from 20 to 80 in steps of 10, with red indicating highest crime and blue indicating lowest.*

Note that the $2 \times 2$ CC map display is a geovisual representation that might be associated with a two-way analysis of variance or with a two-predictor regression model if literacy and wealth are treated as quantitative variables. In a regression of property crime on literacy and wealth, the $R^2$ is a respectable 0.27. However, the standard displays associated with such models seem less useful than CC maps here. For example, Figure 22 shows both the predicted values and the residuals for this model. The predicted values are largely interpretable as showing higher crime in the north as we saw before, but the residuals do not have any obvious interpretations. Moreover, the context provided by the levels of literacy and wealth in the CC map is lost.

## 5. SUMMARY AND CONCLUSIONS

The major goal of this paper was to suggest that André-Michel Guerry deserves greater recognition in the history of statistics and data visualization than he is generally accorded. The 1833 *Essai* broke new ground in thematic cartography and statistical graphics, and established the quantitative study of moral statistics that gave rise to modern social science. His 1864 comparative study of England and France contains graphic displays that rank among the best statistical graphics produced during what I have called the Golden Age of Graphics (Friendly (2007a)); as I suggested, this study contemplated ideas of multivariate explanation well beyond theory and methods available at the time.

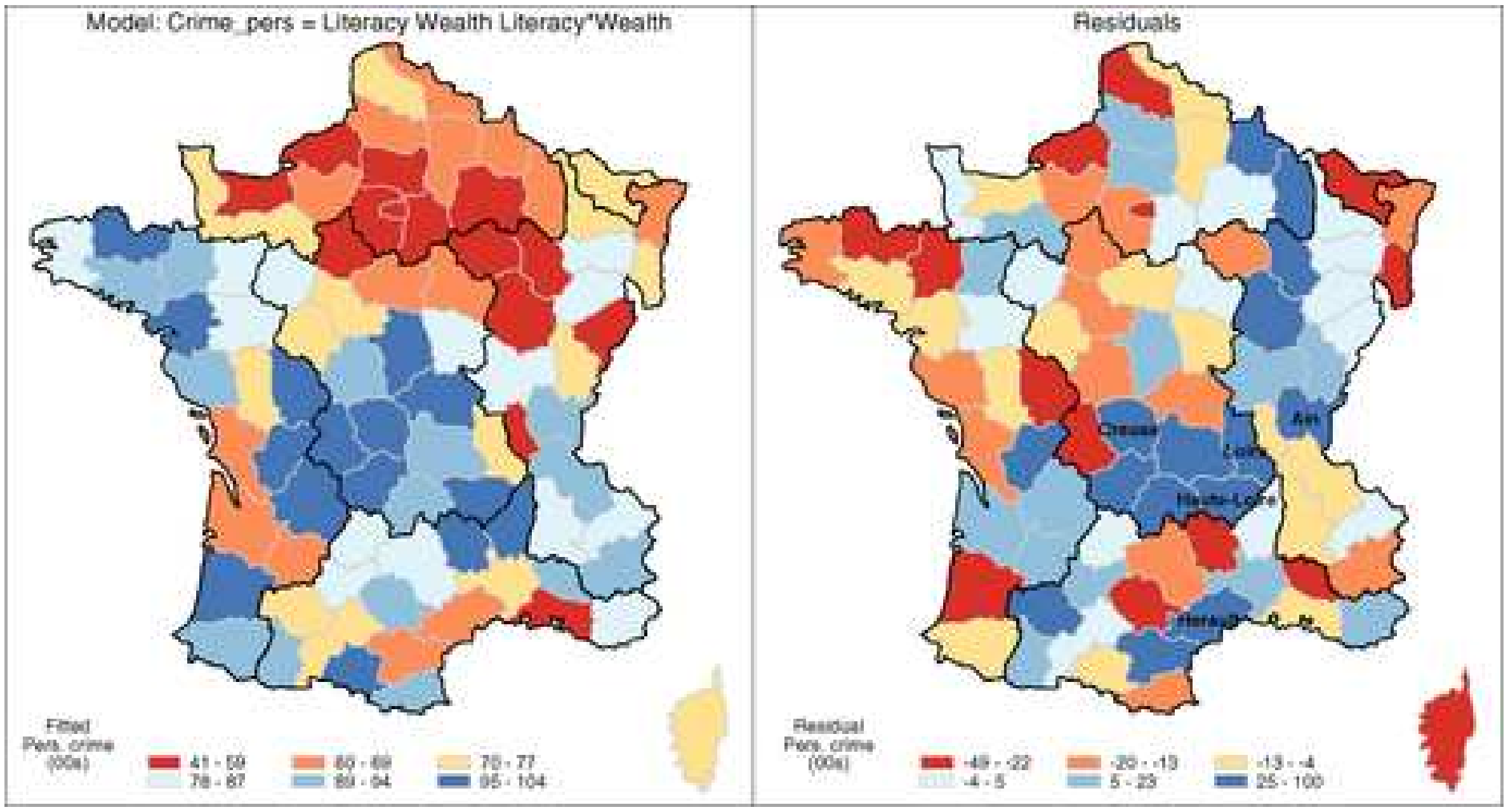

Fig. 22. *Fitted values and residuals from a response surface regression predicting personal crime from literacy and wealth. Départments with "outside" residuals in a boxplot are labeled.*

Yet, as I have also tried to suggest, his questions, methods and data still present challenges for multivariate and spatial visualization today. In particular, what began in Guerry's time as thematic cartography and the first instances of modern statistical maps has progressed to what is now called geovisualization (e.g., Dykes, MacEachren and Kraak (2005)) and exploratory spatial data analysis (ESDA), often providing multiple linked univariate views of geospatial data. Over the same period, statistical analysis developed to encompass multivariate models and graphic displays, but the integration of these data-centric multivariate methods with map-centric visualization and analysis is still incomplete. Who will rise to Guerry's challenge?

## APPENDIX: REGIONS AND DÉPARTEMENTS OF FRANCE IN 1830

As noted earlier, the data from Guerry (1833) used in this article, together with map files in various formats, have been made available at http://www.math.yorku.ca/SCS/Gallery/guerry/. In addition, it may be useful for the reader to have a table listing the names and numbers of the départements of France in 1830, as I have used them in my analyses. Corsica, often an outlier and originally département 20, was subdivided into Haute-Corse (2A) and Corse-du-Sud (2B) in 1975, but is listed in all my files as 200.

## ACKNOWLEDGMENTS

This work is supported by Grant 8150 from the Natural Sciences and Engineering Research Council of Canada. I am grateful to Gilles Palsky, who initiated my interest in Guerry with the gift of images of his maps. Various friends and colleagues, including Jacques Borowczyk, Antoine de Falguerolles, Christian Genet and Ian Spence, helped me correct errors and otherwise strengthen the initial draft, as did the editor and three anonymous reviewers.

| Region | Départment (#) |
|---|---|
| Central | Allier (**3**), Cantal (**15**), Cher (**18**), Corrèze (**19**), Creuse (**23**), Eure-et-Loire (**28**), Indre (**36**), Indre-et-Loire (**37**), Loir-et-Cher (**41**), Loire (**42**), Haute-Loire (**43**), Loiret (**45**), Nièvre (**58**), Puy-de-Dôme (**63**), Sarthe (**72**), Haute-Vienne (**87**), Yonne (**89**) |
| East | Ain (**1**), Basses-Alpes (**4**), Hautes-Alpes (**5**), Aube (**10**), Côte-d'Or (**21**), Doubs (**25**), Drome (**26**), Isère (**38**), Jura (**39**), Haute-Marne (**52**), Meurthe (**54**), Bas-Rhin (**67**), Haut-Rhin (**68**), Rhône (**69**), Haute-Saône (**70**), Saône-et-Loire (**71**), Vosges (**88**) |
| North | Aisne (**2**), Ardennes (**8**), Calvados (**14**), Eure (**27**), Manche (**50**), Marne (**51**), Meuse (**55**), Moselle (**57**), Nord (**59**), Oise (**60**), Orne (**61**), Pas-de-Calais (**62**), Seine (**75**), Seine-Inférieure (**76**), Seine-et-Marne (**77**), Seine-et-Oise (**78**), Somme (**80**) |
| South | Ardèche (**7**), Ariège (**9**), Aude (**11**), Aveyron (**12**), Bouches-du-Rhone (**13**), Gard (**30**), Haute-Garonne (**31**), Gers (**32**), Hérault (**34**), Lot (**46**), Lozère (**48**), Hautes-Pyrénées (**65**), Pyrénées-Orientales (**66**), Tarn (**81**), Tarne-et-Garonne (**82**), Var (**83**), Vaucluse (**84**) |
| West | Charente (**16**), Charente-Inférieure (**17**), Côtes-du-Nord (**22**), Dordogne (**24**), Finestère (**29**), Gironde (**33**), Ille-et-Vilaine (**35**), Landes (**40**), Loire-Inférieure (**44**), Lot-et-Garonne (**47**), Maine-et-Loire (**49**), Mayenne (**53**), Morbihan (**56**), Basses-Pyrénées (**64**), Deux-Sevres (**79**), Vendee (**85**), Vienne (**86**) |
| Other | Corse (**200**) |